\documentclass[lettersize,journal]{IEEEtran}

\IEEEoverridecommandlockouts
\usepackage{algorithm,algorithmic,amsbsy,amsmath,amssymb,epsfig,subfig,bbm,mathrsfs,mathdots,arydshln,arydshln,multirow,verbatim,booktabs,setspace,url,stfloats,cite,nomencl,epstopdf, comment}

\usepackage{xcolor}
\def\BibTeX{{\rm B\kern-.05em{\sc i\kern-.025em b}\kern-.08em
    T\kern-.1667em\lower.7ex\hbox{E}\kern-.125emX}}

\newcommand{\beq}{\begin{equation}}
\newcommand{\eeq}{\end{equation}}
\newcommand{\bitm}{\begin{itemize}}
\newcommand{\ba}{\begin{array}}
\newcommand{\ea}{\end{array}}
\newcommand{\eitm}{\end{itemize}}
\newcommand{\beqn}{\begin{eqnarray}}
\newcommand{\eeqn}{\end{eqnarray}}
\newcommand{\beqno}{\begin{eqnarray*}}
\newcommand{\eeqno}{\end{eqnarray*}}
\newcommand{\bma}{\begin{displaymath}}
\newcommand{\ema}{\end{displaymath}}
\newcommand{\bnu}{\begin{enumerate}}
\newcommand{\enu}{\end{enumerate}}
\newcommand{\bce}{\begin{center}}
\newcommand{\ece}{\end{center}}
\newcommand{\btb}{\begin{tabular}}
\newcommand{\etb}{\end{tabular}}

\begin{document}

%\title{Cooperative Resource Management in Quantum Key Distribution (QKD) for Semantic Communication}
\title{Cooperative Resource Management in Quantum Key Distribution (QKD) Networks for Semantic Communication}
\author{Rakpong~Kaewpuang,
	    Minrui~Xu,
        Wei~Yang~Bryan~Lim,
        Dusit~Niyato,~\IEEEmembership{Fellow,~IEEE},
		Han~Yu,
		\\Jiawen~Kang,
        and Xuemin~(Sherman)~Shen,~\IEEEmembership{Fellow,~IEEE} %<-this % stops a space 
% \IEEEcompsocitemizethanks{\IEEEcompsocthanksitem~Rakpong~Kaewpuang,~and~Dusit~Niyato are with the School of Computer Engineering, Nanyang Technological University~(NTU),~Singapore. \protect \\
% E-mail: jujuve.boo@gmail.com,~\{dniyato\}@ntu.edu.sg \protect\\
%{\bf D. Niyato} is the corresponding author (email: dniyato@ntu.edu.sg). } 
%\IEEEcompsocitemizethanks{ {\bf D. Niyato} is the corresponding author (email: dniyato@ntu.edu.sg). }% <-this % stops a space	
\thanks{Rakpong~Kaewpuang, Minrui~Xu, Dusit~Niyato, and Han~Yu are with the School of Computer Science and Engineering, Nanyang Technological University, Singapore (e-mail: rakpong.kaewpuang@ntu.edu.sg; minrui001@e.ntu.edu.sg; dniyato@ntu.edu.sg; han.yu@ntu.edu.sg).}
\thanks{Wei Yang Bryan Lim is with Alibaba Group and Alibaba-NTU Joint Research Institute, Nanyang Technological University, Singapore (e-mail: limw0201@e.ntu.edu.sg).}
\thanks{Jiawen~Kang is with the School of Automation, Guangdong University of Technology, China (e-mail: kavinkang@gdut.edu.cn).}
\thanks{Xuemin~Sherman~Shen is with the Department of Electrical and Computer Engineering, University of Waterloo, Waterloo, ON, Canada, N2L 3G1 (e-mail: sshen@uwaterloo.ca).}
}
\maketitle

\begin{abstract}
Increasing privacy and security concerns in intelligence-native 6G networks require quantum key distribution-secured semantic information communication (QKD-SIC). In QKD-SIC systems, edge devices connected via quantum channels can efficiently encrypt semantic information from the semantic source, and securely transmit the encrypted semantic information to the semantic destination. In this paper, we consider an efficient resource (i.e., QKD and KM wavelengths) sharing problem to support QKD-SIC systems under the uncertainty of semantic information generated by edge devices. In such a system, QKD service providers offer QKD services with different subscription options to the edge devices. The QKD services are envisioned to follow cloud computing that has the subscription in the reservation and on-demand options, i.e., for long and short (immediate) terms, respectively. As such, to reduce the cost for the edge device users, we propose a QKD resource management framework for the edge devices communicating semantic information. The framework is based on a two-stage stochastic optimization model to achieve optimal QKD deployment. Moreover, to reduce the deployment cost of QKD service providers, QKD resources in the proposed framework can be utilized based on efficient QKD-SIC resource management, including semantic information transmission among edge devices, secret-key provisioning, and cooperation formation among QKD service providers. In detail, the formulated two-stage stochastic optimization model can achieve the optimal QKD-SIC resource deployment while meeting the secret-key requirements for semantic information transmission of edge devices. Moreover, to share the cost of the QKD resource pool among cooperative QKD service providers forming a coalition in a fair and interpretable manner, the proposed framework leverages the concept of Shapley value from cooperative game theory as a solution. Experimental results demonstrate that the proposed framework can reduce the deployment cost by about 40\% compared with existing non-cooperative baselines.
%Consequently, the resources can be utilized and the deployment cost of the QKD service providers can be decreased. To minimize the cost of QKD service providers, we propose resource allocation management for the semantic information transmission of edge devices, cost management, and cooperation formation among QKD service providers. For the resource allocation management to the semantic information transmission, we formulate and solve the two-stage stochastic optimization model to achieve the optimal resources that can be supported to minimize the deployment cost of QKD service providers while satisfying the secret-key requirements for semantic information transmission of edge devices. For sharing the cost of the resource pool among cooperative QKD service providers in a coalition, we apply the concept of Shapley value from cooperative game theory as a solution.
%Based on the cost shares, the QKD service providers can make a decision whether to cooperate and share the resources in the resource pool or not.    

\end{abstract}


\section{Introduction}
\label{sec:introduction}

%QKD
As quantum computers become a distinct possibility, they pose new potent threats to the traditional cryptographic security schemes. This has led  to a revolution in the use of quantum channels for secret key distribution, i.e., quantum key distribution (QKD), for the transmission of confidential information~\cite{cao2022evolution, a-s-cacciapuoti-quantum-2020, a-s-cacciapuoti-entanglement-2020,j-illiano-quantum-internet-2022}. With the parallel processing of quantum computing, quantum computers can easily break the process of key exchanging in modern symmetric cryptography~\cite{ladd2010quantum}. Fortunately, quantum channels based on the laws of quantum mechanics, such as Heisenberg's uncertainty principle and the no-cloning theorem~\cite{wootters1982single}, can perform data transmission, including symmetric key distribution, with provable security. In addition, QKD can enable information-theoretic security (ITS) for data transmission in classical data channels via the one-time pad (OTP) mechanism~\cite{shannon1949communication}. In intelligence-native 6G communications~\cite{you2021towards}, security and privacy are critical requirements in distributed artificial intelligence (AI) training. In particular, the transmission of essential feature information in semantic communication makes 6G communications more demanding for security and privacy in semantic information communication (SIC) and networking.

%semantic communication
Semantic information communication \cite{x-luo-semantic-communications2022, y-wanting-semantic-communi2022, q-zhijin-semantic-comm2022} refers to communication paradigms based on the concept of semantic-meaning transmission. SIC is a practical application of AI models for exchanging meaningful information obtained from transmission data with a minimum of communication resources. The main goal of semantic communication is to extract the meanings or features of the transmitted data (i.e., semantic information) from a semantic source and transmit the semantic information to a semantic destination, where it can be interpreted successfully. In the semantic communication system, we can consider that the semantic source and destination are semantic providers who execute intelligent algorithms (e.g., knowledge graphs, deep neural networks) autonomously. They can be edge devices or sensors with cognitive capabilities. In such a knowledge-based system, the semantic source and destination can establish and utilize their background knowledge bases (KBs) through online or offline training and inference. The KBs are typical models that the semantic source and destination can learn and validate in advance. For example, images, speeches, videos and texts are representative KBs. The semantic source uses an individual KB to extract the semantic information of the raw data, which is then transmitted to the semantic destination. When the semantic destination receives the semantic information, it uses its individual KB to interpret them.

Nevertheless, SIC provides a viable and intelligent way to exchange semantic information between the semantic source and the semantic destination. Regarding security, SIC faces confidentiality issues due to the eavesdropping risk in exchanging semantic information between the semantic source and the semantic target~\cite{j-blesswin-enhanced-semantic2019,m-xu-stochastic-resource2022,kaewpuang2022-adaptive}. To ensure that the transmission of semantic information is secure, QKD is a promising mechanism to protect public keys from eavesdropping attacks.

To address the aforementioned issues, we propose a QKD-secured SIC (QKD-SIC) system to protect the semantic information and public keys from eavesdropping attacks. We consider the resource management problem that the secret keys are allocated to support the semantic information of edge devices in SIC. In addition, we consider that the amount of semantic information generated by edge devices is unpredictable. To optimize QKD resources when the amount of semantic information is unpredictable, QKD service providers can cooperate and establish a global resource pool. In particular, multiple QKD service providers can share their resources (i.e., QKD and KM wavelengths) in the pool. Therefore, the resources in the pool can be efficiently utilized by the QKD service providers. Nevertheless, three important questions arise in the context of this resource management. Firstly, what is an optimal resource allocation in the pool of SIC applications to minimize costs and satisfy the needs of edge devices? Secondly, how can the costs incurred by the resource pool be shared among cooperative QKD service providers? Thirdly, should QKD service providers cooperate or not in creating the resource pool? To answer these questions for QKD service providers, we introduce a decision-making scheme with the components of resource allocation to SIC applications, cost management, and collaboration among QKD service providers in a QKD-SIC environment. The goal of the scheme is to make the best decisions for QKD service providers as they are rational and interested in minimizing their own costs. The main contributions of this paper can be summarized as follows:
\begin{itemize}
    \item We introduce a hierarchical architecture for the QKD-SIC system, where resource allocation and routing decisions are coupled. To provide sufficient secret keys for semantic information transmission between virtual service providers and edge devices, the QKD network manager and controller allocate QKD resources (i.e., QKD and KM wavelengths) from QKD nodes and determine routing paths for QKD links. 
    \item We formulate and solve the stochastic programming (SP) model to obtain the optimal resource allocation decisions from a resource pool created by QKD service providers. The SP model requires the probability distributions of the random parameters, i.e., the secret key rate (user demand) requirements. This model can be applied for the cooperative QKD service providers to make decisions in two stages. In the first stage, the QKD service providers make a resource reservation decision (i.e., QKD and KM wavelengths) based on the statistical information about users' demand. In the second stage, when QKD service providers know the exact level of users' demand, they make a decision to compensate for users' demand that cannot be satisfied when resources are insufficient. 
    \item We present a model to allocate the provisioning cost obtained from the resource pool among QKD service providers in a fair manner. In addition, we leverage Shapley values to determine fair and interpretable allocation of provisioning costs. 
%    \item We propose a game model for the formation of cooperation among QKD service providers to decide whether they should cooperate and create the resource pool or not. 
    \item We introduce a cooperative game model for QKD service providers to decide whether they should cooperate and establish the resource pool or not. A stable cooperation strategy is the solution of the game model, which means that the rational QKD service providers do not want to change their decisions.  
\end{itemize}

The remainder of this paper is organized as follows: Section~\ref{sec:related-workd} presents related work. Section~\ref{sec:system-model} describes the system model, the maritime case study, and the cost models. Section~\ref{sec:optimization-formulation} presents the stochastic programming model for resource allocation in QKD-SIC for a maritime application. Section~\ref{sec:cost-management} presents cost management model among QKD service providers. Section~\ref{sec:cooperation-formation} presents the cooperative game models for optimal cooperation formation for QKD service providers. Section~\ref{sec:performance-evaluation} presents the performance evaluation results. Section~\ref{sec:conclusion} concludes the paper.

%The remainder of this paper is organized as follows: Section~\ref{sec:related-workd} presents related work. Section~\ref{sec:system-model} describes the system model, the maritime case study, and the cost models. Section~\ref{sec:optimization-formulation} presents the stochastic programming model for resource allocation in QKD-SIC for a maritime application. Section~\ref{sec:cost-management} presents the model for cost management among QKD service providers. Section~\ref{sec:cooperation-formation} presents the game models for optimal cooperation formation for QKD service providers. Section~\ref{sec:performance-evaluation} presents the performance evaluation results. Section~\ref{sec:conclusion} concludes the paper.

\section{Related Work}
\label{sec:related-workd}

\subsection{Semantic Information Communication}
Semantic information communication significantly reduces the need for transmission resources such as channels by using artificial intelligence to extract semantic features for the information between source and destination~\cite{zaarour2022openpubsub, kim2020compiler,hu2019things2vec,qiu2020mobile}.
The authors in \cite{w-c-ng-stochastic-approach2022} proposed a SIC in the metaverse \cite{n-h-chu-metaslicing2022} between edge devices and the virtual service providers (VSPs) to reduce the size of transmitted data. In SIC, an edge device (e.g., a smartphone) can autonomously produce pictures (pre-semantic data) relevant to the interest of VSPs. The VSP can subscribe to edge devices that the VSP is interested in transmitting the semantic data to create the metaverse for supporting user requests, where the user requests are considered as uncertainty. In \cite{w-c-ng-stochastic-approach2022}, the stochastic semantic transmission scheme was proposed to minimize the transmission cost of VSPs. This scheme was formulated as a two-stage stochastic integer programming model under the uncertainty of user requests to achieve an optimal solution. The resource allocation for text SIC was proposed in \cite{yan-lei-resource-allocation2022} to maximize the overall semantic spectral efficiency in terms of channel utilization and the number of transmitted semantic symbols of all users. The authors in \cite{yan-lei-resource-allocation2022} compared the optimal solution of the proposed scheme with that of three benchmarks. The simulation results demonstrated that the performance of the proposed scheme outperformed that of the three benchmarks. The deep learning-enabled SIC system for speech signals (DeepSC-S) was proposed in \cite{w-zhenzi-semantic-comm2021} to enhance the recovery accuracy of speech signals. Particularly, the joint semantic-channel coding was proposed to solve source distortions and channel effects. The squeeze-and-excitation network was employed in the proposed DeepSC-S to learn and extract speech semantic information. In experimental results, the performance of the proposed DeepSC-S was compared with the traditional communication system and the system with an additional feature coding for speech transmission. In addition, the proposed DeepSC-S was applied to telephone systems and multimedia transmission systems. However, current SIC systems only leverage AI models to transmit necessary semantic information between the source encoder and the destination decoder, while the security issues in SIC systems are largely overlooked.

\subsection{The Quantum Internet}

With the ability to provide information-theoretic security for communications, QKD and QKD-secured systems are expected to protect various 6G communications such as optical, satellite, maritime, and their cross-layer communications~\cite{ma2022equilibrium,zhang2020flexible, you2021towards, m-xu-quantum-secured-space2022}. Unlike traditional QKD protocols, such as Bennett-Brassard-1984 (BB84) \cite{b-chales-h-2014-quantum-cryp}, Bennett-Brassard-Mermin-1992 (BBM92)~\cite{cao2022evolution}, and Grosshans-Grangier-2002 (GG02)~\cite{cao2022evolution}, modern QKD protocols based on MDI technology can provide longer distribution distance and stronger security without assuming trusted relays~\cite{lo2012measurement}. To minimize the cost of deploying QKD resources in MDI-QKD, the authors in~\cite{y-cao-hybrid-trusted2021} proposed a static linear programming model and the CO-QBN algorithm to efficiently manage QKD networks. The authors in \cite{kaewpuang2022resource} proposed the resource allocation scheme for quantum-secured space-air-ground integrated networks (SAGIN) in which QKD services protect secure communications between space, aerial, and ground nodes by exchanging secret keys in quantum channels. In \cite{kaewpuang2022resource}, the authors formulated and solved the stochastic programming model to achieve the optimal solution for resource allocation (i.e., QKD and KM wavelengths) and routing under the uncertainties of secret-key rate requirements and weather conditions. 

In \cite{m-xu-stochastic-resource2022}, the authors proposed the QKD-based secure federated learning (FL) scheme to support the FL model encryption against eavesdropping attacks on FL networks. Particularly, the resource allocation scheme for QKD resources (i.e., wavelengths) to support FL networks was proposed to minimize the overall deployment cost under the uncertainty of the number of FL workers associated with the secret-key rate requirements. The resource allocation scheme was formulated based on the two-stage SP to achieve the optimal solution under the uncertainty of the number of FL workers. The concept of QKD over SAGIN was proposed to achieve secure communications with global coverage and reconfigurable \cite{h-cui-space-air-ground2022}, \cite{m-xu-quantum-secured-space2022}. The authors \cite{m-xu-quantum-secured-space2022} proposed the QKD service provisioning framework to establish secure communications in quantum-secured SAGIN. The two-stage SP was formulated to minimize the provisioning cost under the uncertainty of secret-key rates. In practice, the framework was applied to support metaverse applications to achieve the optimal solution. The authors \cite{kaewpuang2022-adaptive} proposed the hierarchical architecture for QKD-secured federated learning systems. In \cite{kaewpuang2022-adaptive}, the QKD resource management was formulated, which is based on the two-stage SP, to minimize the deployment cost under the uncertainty of the secret-key rate requirements from FL applications. \cite{s-k-liao-satellite-to-ground2017}, \cite{j-yin-entanglement-based2020} introduced systems to provide QKD services via free space to secure applications to satisfy security requirements. The authors in \cite{j-yin-entanglement-based2020} demonstrated the quantum satellite (i.e., Micius) that successfully distributes the secret key (i.e., with the rate of 0.12 bits per second) between two ground nodes via free space. The drone-based entanglement distributions were proposed in \cite{h-y-liu-optical-relayed2021}, \cite{h-y-liu-drone-based-entanglement2020}. In \cite{h-y-liu-drone-based-entanglement2020}, the mobile quantum communications via entanglement distribution demonstrated that the proposed system can tolerate various weather conditions (i.e., daytime, clear nights, and rainy nights). Compared to fiber- and satellite-based quantum communications, drone-based entanglement distribution offers the advantages of unparalleled mobility, flexibility, and reconfigurability.

\subsection{Game-based Resource Allocation}

The authors in \cite{g-basak-semantic-comm-game2018} proposed a game-based resource allocation framework for SIC system, where the semantic content of a message to be transmitted over a noisy channel was taken into consideration to optimize the performance of a communication system. In addition, the impact of social influence on how the messages were translated by considering the influence of the agent in the communication network. The agent had an influence on the receiver, i.e., how to decode the received messages by providing side information. The influence of the agent can be adversarial or helpful to the communicating parties. Therefore, the authors considered the SIC problem with social influence as a Bayesian game with incomplete information and investigated the conditions under which a Bayesian Nash equilibrium exists.

Nevertheless, the existing works in the literature do not consider the uncertainty of secret-key requirements from SIC. In addition, the works in the literature do not consider the problem of optimizing the resource allocation to semantic applications, cost management, and cooperation formation for QKD service providers under uncertainty.

%Nevertheless, the existing works on resource management in the literature do not consider the uncertainty of secret-key requirements from SIC. In addition, none of these works in the literature considers the problem of jointly optimizing the resource allocation to semantic applications, cost management, and cooperation formation for QKD service providers under uncertainty.

\section{System Model and Case Study}
\label{sec:system-model}

\begin{figure}[htb]
\centering
\captionsetup{justification=centering}
$\begin{array}{c} \epsfxsize=3.3in \epsffile{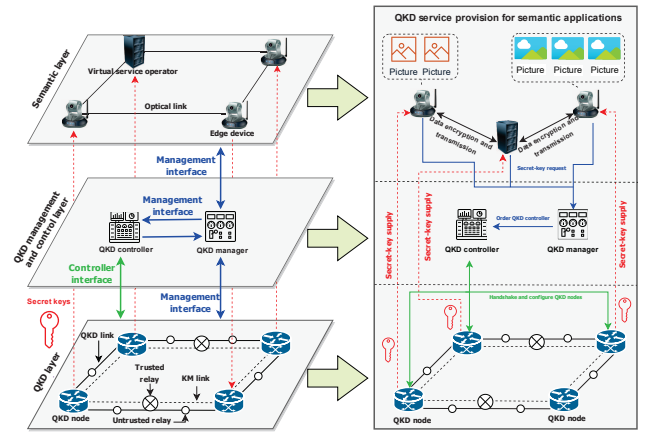} \\
\end{array}$
\caption{The QKD-SIC architecture.} 
\label{fig:system-model}
\end{figure}

We propose a three-layer QKD-SIC architecture illustrated in Fig. \ref{fig:system-model}. The architecture consists of the QKD layer, the QKD management and control layer, and the semantic layer. In the QKD layer, there are three types of nodes and two types of links. The three types of nodes are QKD nodes, trusted relays, and untrusted relays. The two types of links are key management (KM) links and QKD links. We assume that QKD nodes are co-located with SIC nodes, which can be edge devices or quantum secure link users. The links (i.e., QKD, KM, and optical links) are multiplexed within a single fiber by MUX/DEMUX components \cite{y-cao-hybrid-trusted2021}. Therefore, the network topology for the QKD layer is the same as that for the semantic layer. Let $\mathcal{N}$ and $\mathcal{E}$ denote the set of SIC/QKD nodes and the set of fiber links between SIC/QKD nodes, respectively. 

A QKD node supplies secret keys to its co-located SIC node. The trusted relay and the untrusted relay are located between the QKD nodes. The QKD node consists of a global key server (GKS), a local key manager (LKM), a security infrastructure (SI), and components of trusted/untrusted relays. A trusted relay consists of two or multiple transmitters of MDI-QKD (MDI-Txs), LKM, and SI, while an untrusted relay consists of one or multiple receivers of MDI-QKD (MDI-Rxs). An MDI-QRx must be located between two MDI-QTxs to generate local secret keys. An LKM is used to obtain and store the local secret keys from the connected MDI-QTxs, and LKM performs secret key propagation via the one-time-pad (OTP) method \cite{y-cao-hybrid-trusted2021} to generate global secret keys between QKD nodes. The SIs are applied to ensure that trusted relays work in a secure manner. In the QKD layer, the QKD links are used to connect between MDI-QTxs and MDI-QRxs, while the KM links are used to connect between LKMs. The QKD link contains quantum and classical channels, while the KM link contains a classical channel implemented by wavelengths. 

In the semantic layer, an SIC node can be an edge device (e.g., a smartphone and surveillance camera) or a virtual service operator (i.e., semantic receiver). Each edge device in different areas has different views for locations, as shown in Fig. \ref{fig:system-model}. The edge device can capture the pictures and extract the semantic information (i.e., snapshots) from the pictures. Then, the edge device transmits the semantic information via optical fibers to the respective virtual service operator (VSO), which uses this data to create the virtual environment for a particular application, for example, transportation and education. Before the semantic information transmission between the edge device and the VSO happens, the VSO has to subscribe to the respective edge devices.

When the VSO subscribes to the respective edge devices, the respective edge devices will request secret keys from the QKD manager based on their security requirements for encrypting their semantic information. When the QKD manager receives the request, the QKD manager responds by checking the available secret keys. Suppose the secret keys are available to support the request, and thus the QKD manager will order the QKD controller to select the suitable QKD nodes to provide the secret keys in a suitable format for semantic information encryption. If the secret keys are unavailable to support the request, the edge devices and the VSO will wait until the secret keys are available. Then, the edge devices encrypt the semantic information via the secret keys and transmit the encrypted semantic information to the semantic receiver. The amount of semantic information (i.e., snapshots) of each edge device directly depends on the available secret-key resources in the QKD layer as well as the security requests in the semantic layer. Therefore, it has a trade-off between the QKD resources and the requests from edge devices. 

\subsection{Cooperation Among QKD Service Providers and Pooling of QKD Resources}
\label{subsec:cooperation-among-qkd-service-providers}

\begin{figure}[htb]
\centering
\captionsetup{justification=centering}
$\begin{array}{c} \epsfxsize=3 in \epsffile{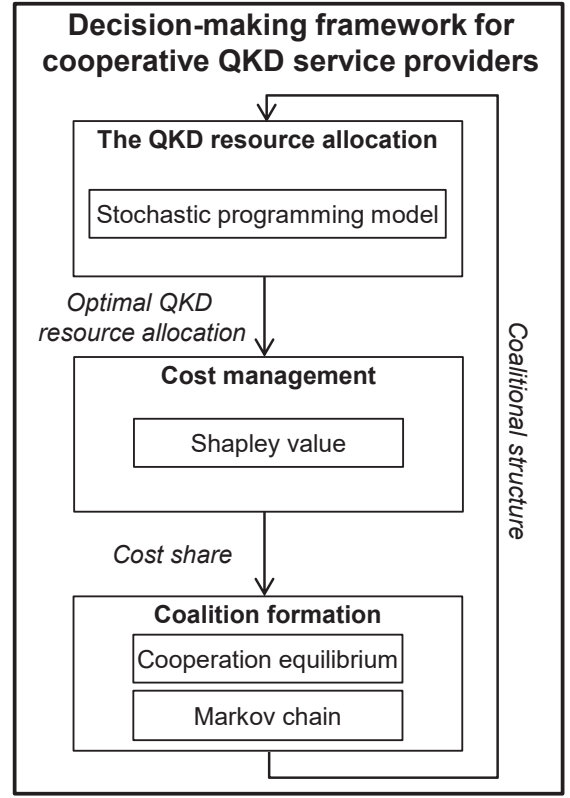} \\
\end{array}$
\caption{Components of the decision-making framework for QKD service providers.} 
\label{fig:framework}
\end{figure}

To minimize the deployment costs, the QKD service providers can cooperate and establish a QKD resource pool that comprises QKD and KM wavelengths to support the security demand. The set of cooperative QKD service providers that agree to cooperate and establish the QKD resource pool is represented by $\mathbb{C}$. It could have many coalitions. Therefore, let $\Upsilon = \{ {\mathbb{C}}_1,{\mathbb{C}}_2,{\mathbb{C}}_3,\ldots, {\mathbb{C}}_{|\Upsilon|}\}$ denote a set of all coalitions (i.e., cooperation structure or coalitional structure) where ${\mathbb{C}}_i \in \Upsilon$. Each coalition $i$ ($\mathbb{C}_{i}$) has its own QKD resource pool that is created by the cooperative QKD service providers in the coalition $i$. Given a coalition ${\mathbb{C}}$, $W^{\mathrm{qkd}}_{i,j}(\mathbb{C})$ denotes available QKD wavelengths on the QKD link between SIC/QKD nodes $i$ and $j$ in the corresponding resource pool. $W^{\mathrm{kml}}_{i,j}(\mathbb{C})$ denotes available KM wavelengths on the KM link between SIC/QKD nodes $i$ and $j$ in the corresponding resource pool. The cost generated from the QKD resource pool will be shared by all cooperative QKD service providers in the coalition.

With the collaboration among the QKD service providers, we introduce a framework that comprises three main components.
\begin{itemize}
\item { \emph{QKD resource allocation for QKD service providers:}} With cooperative QKD service providers, the objective of QKD resource allocation is to procure the optimal numbers of QKD and KM wavelengths while the deployment cost of the cooperative QKD service providers is minimized. Hence, we formulate an optimization problem based on SP. The probability distributions of random parameters in the SP model are taken into consideration. 
\item {\emph{Cost management:}} Given a coalition of QKD service providers and the optimal QKD resource allocation, we apply the concept of Shapley value to obtain the share of deployment cost for each cooperative QKD service providers. 
\item {\emph{Coalition formation:}} We analyze the stable cooperation strategy among QKD service providers by using the cooperation formation model. The dynamics of coalition formation are analyzed by Markov chain models.
\end{itemize}

Figure~\ref{fig:framework} illustrates the interactions among the three components. The coalition formation is first performed to achieve the coalition structure and then the coalition structure is applied by the QKD resource allocation. The optimal QKD resource allocation of the SP model is used by the cost management in which the Shapley value is applied to allocate the incurred deployment cost to QKD service providers. The QKD service providers alter their coalition formation strategies accordingly when they obtain their costs. 

\subsection{The Network Model}
\label{subsec:network_model}
In the QKD-SIC model, let $\mathcal{F}$ and $f(s_{f}, d_{f}, P_{f}(\cdot)) \in \mathcal{F}$ denote the set of QKD-SIC chain requests and a QKD-SIC chain request of SIC nodes, respectively. $s_{f}$ and $d_{f}$ are the source node and destination node of QKD-SIC chain request $f$, respectively. Let $K_{D}$ denote the maximum achievable secret-key rate at a distance $D$. The number of parallel QKD-SIC links $P_{f}(\cdot)$ to satisfy the security demand (secret-key rate) of SIC transmission between $s_{f}$ and $d_{f}$ is expressed as follows: 
\begin{equation}
	P_{f}(\tilde{\omega}) = \Big\lceil \frac{\tilde{\kappa}_{f}}{K_{D}}	\Big\rceil \label{eq:parallel-flqkd-links}		
\end{equation}
where $f$ is the QKD-SIC chain request and $\tilde{\kappa}_{f}$ is a random variable of secret-key rate requirement of request $f$. $D$ is a distance between two connected MDI-QTxs, which can be expressed in as follows:  
\begin{equation}
	D \approx 2 \phi \label{eq:distance_mdi-qtxs} 			
\end{equation}
where $\phi$ is the distance between the MDI-QRx connecting to the MDI-Tx. 

\subsection{The Cost Model}
\label{subsec:cost_model}

We adopt the costs of the QKD network components and the links from \cite{y-cao-hybrid-trusted2021} to support the deployment of QKD with hybrid trusted/untrusted relays on the existing optical backbone network. The QKD network components include MDI-QTxs, MDI-QRxs, LKMs, SIs, and multiplexing/demultiplexing (MUX/DEMUX) components. The links include QKD links and KM links. The components and links can be described as follows: 

\subsubsection{MDI-QTxs and MDI-QRxs} The MDI-QKD process requires two MDI-QTxs and one MDI-QRx and therefore the number of MDI-QTxs ($A^{f}_{\mathrm{tx}}(\cdot)$) and MDI-QRxs ($A^{f}_{\mathrm{rx}}(\cdot)$), which satisfy the QKD-SIC chain request $f$, are expressed in (\ref{eq:cost-mdi-qtxs}) and (\ref{eq:cost-mdi-qrxs}), respectively. 
\beqn
		A^{f}_{\mathrm{tx}} (\tilde{\omega}) = \sum_{i \in \mathcal{N}_{f}} \sum_{j \in \mathcal{N}_{f}} 2 P_{f}(\tilde{\omega}) \Big\lceil \frac{e_{(i,j)}} {D}	\Big\rceil   \label{eq:cost-mdi-qtxs} \\
		A^{f}_{\mathrm{rx}} (\tilde{\omega}) = \sum_{i \in \mathcal{N}_{f}} \sum_{j \in \mathcal{N}_{f}}   P_{f}(\tilde{\omega}) \Big\lceil \frac{e_{(i,j)}} {D}	\Big\rceil   \label{eq:cost-mdi-qrxs}
\eeqn
$e_{(i,j)}$ is the distance of physical fiber link between nodes $i$ and $j$. $\mathcal{N}_{f}$ is set of nodes having the fiber links on the route of request $f$. 

\subsubsection{Local key managers (LKMs)} For the QKD-SIC chain request $f$, the required number of LKMs ($A^{f}_{\mathrm{km}}$) is expressed as follows:
\beqn
	   A^{f}_{\mathrm{km}} =  \sum_{i \in \mathcal{N}_{f}} \sum_{j \in \mathcal{N}_{f}} \Big\lceil \frac{e_{(i,j)}} {D} + 1 \Big\rceil.   \label{eq:cost-lkms} 
\eeqn

\subsubsection{Security Infrastructures (SI)} The required number of SIs ($A^{f}_{\mathrm{si}}$) to satisfy the QKD-SIC chain request $f$ is expressed as follows:
\beqn
	   A^{f}_{\mathrm{si}} =  \sum_{i \in \mathcal{N}_{f}} \sum_{j \in \mathcal{N}_{f}} \Big\lceil \frac{e_{(i,j)}} {D} - 1 \Big\rceil.   \label{eq:cost-si} 
\eeqn
\subsubsection{MUX/DEMUX components} The required number of MUX/DEMUX component pairs ($A^{f}_{\mathrm{md}}$) for the QKD-SIC chain request $f$ is expressed as follows:
\beqn
	A^{f}_{\mathrm{md}} =  \sum_{i \in \mathcal{N}_{f}} \sum_{j \in \mathcal{N}_{f}} \Big\lceil \frac{e_{(i,j)}} {D} \Big\rceil  +   \sum_{i \in \mathcal{N}_{f}} \sum_{j \in \mathcal{N}_{f}} \Big\lceil \frac{e_{(i,j)}} {D} - 1 \Big\rceil.  \label{eq:cost-mux-demux}
\eeqn

\subsubsection{QKD and KM links} Three wavelengths and one wavelength are occupied by the QKD link and the KM link, respectively \cite{a-wonfor-field-trail2019}. The link cost for the QKD-SIC chain request $f$ is expressed as follows:  
\beqn
	A^{f}_{\mathrm{ch}} ( \tilde{\omega} ) =  \sum_{i \in \mathcal{N}_{f}} \sum_{j \in \mathcal{N}_{f}} ( 3 P_{f}(\tilde{\omega}) e_{(i,j)} + e_{(i,j)} )  \label{eq:cost-qkd-km-link} 
\eeqn
where the physical lengths of QKD links and KM links are denoted by $3P_{f}(\tilde{\omega}) e_{(i,j)}$ and $e_{(i,j)}$, respectively.

\subsection{The Cooperation Cost}
\label{subsec:the_cooperation_cost}

We consider the cooperation cost of QKD service providers when the providers decide to share the resources (i.e., QKD and KM wavelengths) in the resource pool. The costs of QKD and KM wavelengths to be shared in the pool can be expressed in (\ref{eq:qkd-wavelength-sharing-cost}) and (\ref{eq:km-wavelength-sharing-cost}), respectively.  
\beqn
	 & & C^{\mathrm{qkd}}(s) = W^{\mathrm{qkd}}_{s} C^{\mathrm{qkd}}_{s} \label{eq:qkd-wavelength-sharing-cost} \\
	 & & C^{\mathrm{kmw}}(s) = W^{\mathrm{kmw}}_{s} C^{\mathrm{kmw}}_{s}  \label{eq:km-wavelength-sharing-cost} 
\eeqn
$W^{\mathrm{qkd}}_{s}$ and $W^{\mathrm{kmw}}_{s}$ are the number of QKD wavelengths and the number of KM shared in the pool by QKD service provider $s$, respectively. $C^{\mathrm{qkd}}_{s}$ and $C^{\mathrm{kmw}}_{s}$ are the costs of shared QKD and KM wavelengths to be charged by QKD service provider $s$, respectively. In addition, let $C^{\mathrm{coc}}_s$ denote the fixed cooperation cost that will be charged when QKD service provider $s$ is in cooperation. This cost can incur due to the communication and computation overhead and other security management tasks, which is typically incurred when independent providers interact with each other.  

\subsection{The Case Study: Maritime Transportation Surveillance over QKD}
\label{subsec:a-case-study}

In the case study of maritime transportation surveillance in QKD-SIC, we give an overview of maritime transportation and highlight the semantic requirements (i.e., semantic information) of maritime transportation applications.

\begin{figure*}[htb]
\centering
\captionsetup{justification=centering}
$\begin{array}{c} \epsfxsize=7in \epsffile{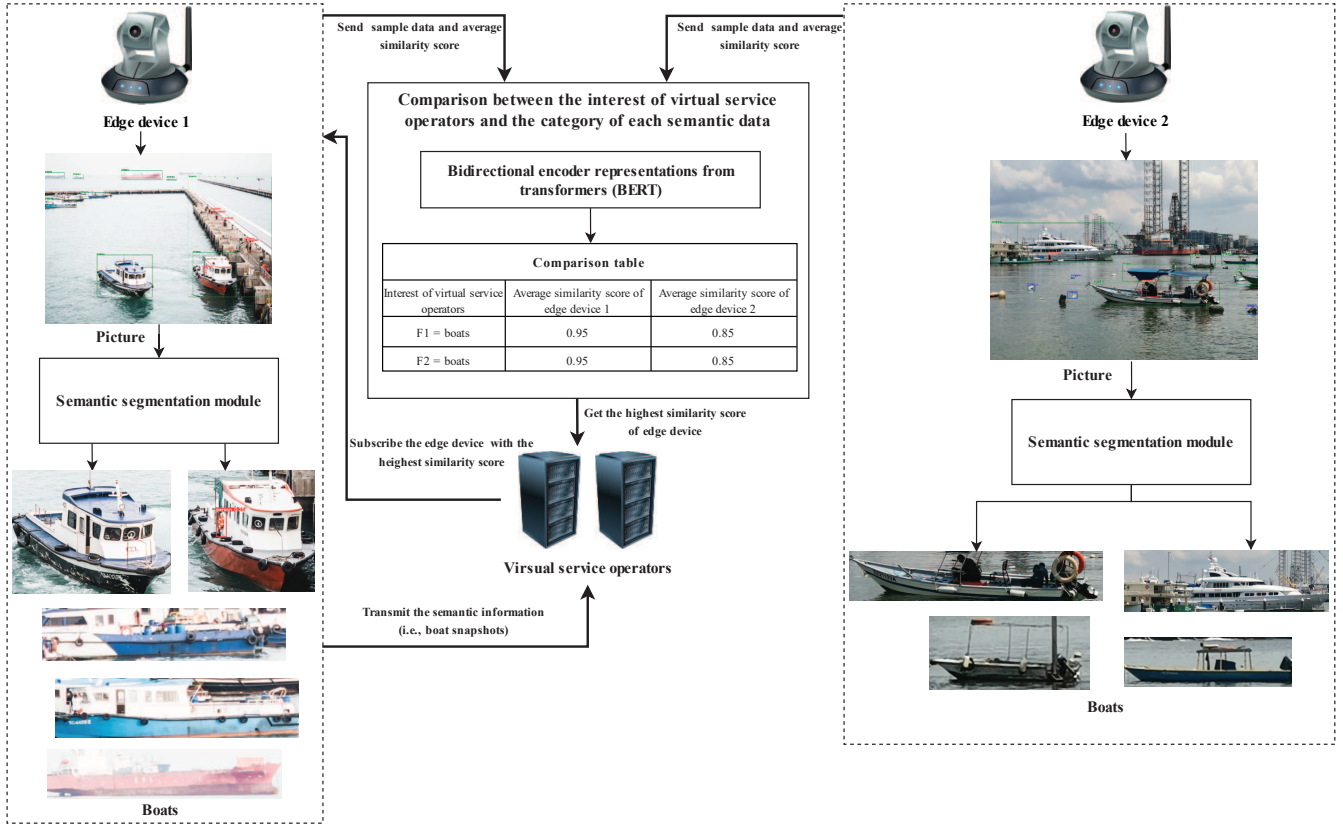} \\
\end{array}$
\caption{An example of a maritime transportation surveillance semantic information transfer.} 
\label{fig:maritime-transportation}
\end{figure*}

Maritime transportation, where goods or people are transported via sea routes, is introduced as a case study. In this example, we consider maritime transportation from the perspective of a virtual service operator (VSO), i.e., semantic receiver, who participates in the virtual environment~\cite{x-luo-semantic-communications2022}, \cite{n-h-chu-metaslicing2022}. It is challenging for new employees to practically sail ships since they are unfamiliar with locations and procedures. This may cause accidents with employees and boatmen. Therefore, a virtual environment for maritime application is a promising solution. The VSO can create a virtual environment of maritime applications for practicing employees.

In Fig. \ref{fig:maritime-transportation}, there are two VSOs and two edge devices in the system. We consider that the VSOs are maritime shipping companies to provide the services of autonomous maritime ships. The VSO can set up simulated environments in the virtual environment in which the VSO can deploy their employee to use the virtual environment and experiment with the autonomous maritime ships. To create the maritime virtual environment, the VSO has to subscribe to edge devices to request the semantic information. Before subscribing to edge devices, the VSO uses bidirectional encoder representations from transformers (BERT) \cite{h-xie-deep-learning2021} with the average similarity score of the category of each semantic information that generated by edge devices. For example, in Fig. \ref{fig:maritime-transportation}, edge devices send the sample data and the average category similarity scores to VSOs. When VSOs receive the sample data and the average category similarity scores, VSOs extract the semantic information (i.e., snapshots of boats) in categories and then compare their interest (i.e., boats) with the similarity scores. As shown in Fig. \ref{fig:maritime-transportation},  the average similarity score of semantic information of edge device 1 is the highest score (i.e., 0.95). Therefore VSOs subscribe edge device 1 to transmit the semantic information.

The VSO can create and provide a real-time virtual environment for maritime applications to support its users. The virtual environment of maritime applications can interact with edge devices by transmitting semantic information back and forth since the VSO can build the details of the virtual environment. The edge devices can be considered to be surveillance cameras or smartphones that are located in different areas. The edge devices and VSO are connected with the optical fibers. The edge devices generate the semantic information by using the semantic segmentation modules. In the semantic segmentation module, the pre-trained machine learning models are applied, e.g., you only look once (YOLO) \cite{j-redmon-computer-vision2018}. The snapshots from edge devices are considered to be the semantic information of interest to the VSO. When the VSO receives the semantic information, the VSO uses this semantic information to build the virtual environment of maritime applications for users.

Therefore, to prevent eavesdropping attacks, the QKD is promising for providing proven secure key distribution schemes for the semantic information transmitting between edge devices and the VSO by facilitating public key and data encryption. In this paper, we consider that the number of secret-key rate requirements in the QKD layer directly depends on the number of snapshots captured by edge devices. However, the number of snapshots of each edge device can not be known in advance since the number of snapshots which are generated by the semantic segmentation module is relevant to the interest of VSO. Therefore, the number of snapshots of each device becomes an uncertain parameter.

\begin{figure}[htb]
\centering
\captionsetup{justification=centering}
$\begin{array}{c} \epsfxsize=3 in \epsffile{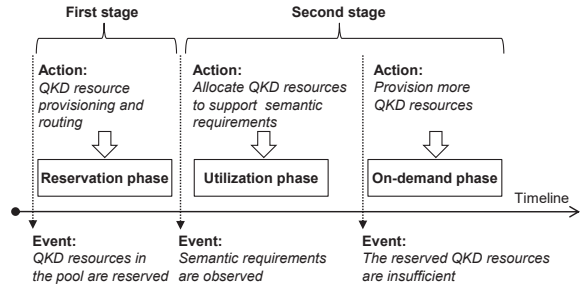} \\
\end{array}$
\caption{The two-stage stochastic programming.} 
\label{fig:two-stage-sp}
\end{figure}

The QKD resource allocation for maritime transportation applications from the QKD resource pool that is created by QKD service providers in coalition $\mathbb{C}$ is formulated as the two stages of the SP model. Figure~\ref{fig:two-stage-sp} illustrates the two stages of the SP model for maritime transportation applications given the QKD resource pool that is conceived by cooperative QKD service providers in coalition $\mathbb{C}$. In the first stage, decisions of provisioning QKD and KM wavelengths, and a decision of the routing according to the semantic requirements (i.e., from the semantic source node to destination node) are made. In the second stage, when the semantic requirements (i.e., the number of snapshots generated by an edge device) are observed, the QKD and KM wavelengths that are provisioned in the first stage will be allocated to satisfy the semantic requirements. If the numbers of reserved QKD and KM wavelengths in the first stage are not enough, the QKD and KM wavelengths in the on-demand phase are provisioned.  

\section{Optimization Formulation}
\label{sec:optimization-formulation}

In this section, we first describe sets, constants, and decision variables of the SP model. We next present the SP model from the QKD resource pool created by the cooperative QKD service providers in coalition $\mathbb{C}$. Finally, we present the deterministic equivalent formulation to achieve the solution of the SP model.

\subsection{Model Description}
\label{subsec:model-description}
The optimization model is based on the two-stage SP \cite{Brige1997}. The sets and constants in the optimization are defined as follows: 
\begin{itemize}
	\item ${\mathcal{N}}$ denotes a set of all nodes in the network.
	\item ${\mathcal{O}}_n$ denotes a set of outgoing links from node $n \in {\mathcal{N}}$.
	\item ${\mathcal{I}}_n$ denotes a set of incoming links to node $n \in {\mathcal{N}}$.
	\item ${\mathcal{F}}$ denotes a set of QKD-SIC chain requests in the network.
%	\item $W^{\mathrm{qkd}}_{i,j}(\mathbb{C})$ denotes available QKD wavelengths on the QKD link between SIC/QKD nodes $i$ and $j$ ($i, j \in \mathcal{N}$), which is created by QKD service providers in coalition structure $\mathbb{C}$.
%	\item $W^{\mathrm{kml}}_{i,j}(\mathbb{C})$ denotes available KM wavelengths on the KM link between SIC/QKD nodes $i$ and $j$, which is created by QKD service providers in coalition structure $\mathbb{C}$.
	\item $W^{\mathrm{qkd}}_{f}$ denotes 3 wavelengths of QKD link of $f \in {\mathcal{F}}$ satisfying the secrete-key rate~\cite{y-cao-hybrid-trusted2021}.
	\item $W^{\mathrm{kml}}_{f}$ denotes 1 wavelength of KM link of $f \in {\mathcal{F}}$ satisfying the secrete-key rate~\cite{y-cao-hybrid-trusted2021}.
	\item $B^{\mathrm{eng}}_{n, f}$ denotes the amount of energy required for transferring traffic of request $f \in {\mathcal{F}}$ through node $n \in {\mathcal{N}}$.
\end{itemize}
We denote $S_f \in {\mathcal{N}}$ and $D_f \in {\mathcal{N}}$ as the source node and the destination node of request $f$, respectively. The decision variables of the network are as follows:
\begin{itemize}
	\item $x_{i,j,f}$ is a binary variable indicating whether request $f \in {\mathcal{F}}$ will take a route with the link from node $i \in {\mathcal{N}}$ to node $j \in {\mathcal{N}}$ or not.
	\item $y^{\mathrm{r}}_{i,n,f}$ and $z^{\mathrm{r}}_{i,n,f}$  are non-negative variables indicating the wavelength channels for QKD and KM links in the reservation phases, respectively.
	\item  $y^{\mathrm{e}}_{i,n,f}$ and $z^{\mathrm{e}}_{i,n,f}$ are non-negative variables indicating the utilized wavelength channels for QKD and KM links, respectively.
	\item  $y^{\mathrm{o}}_{i,n,f}$ and $z^{\mathrm{o}}_{i,n,f}$ are non-negative variables indicating the wavelength channels for QKD and KM links in the on-demand phases, respectively. 
\end{itemize}

\subsection{Stochastic Programming Formulation}
\label{subsec:stochastic-programming}

We first propose a QKD resource allocation scheme based on a two-stage SP model among cooperative QKD service providers in coalition structure $\mathbb{C}$. The resource allocation decisions in the proposed scheme are conducted in two stages according to the result of the SP model.

%%%%%%%%%%%%%%% Stochastic version %%%%%%%%%%%%%%%%%%%%
\beqn
	v (\mathbb{C}) & & = \min  \sum_{f \in {\mathcal{F}}} \sum_{n \in {\mathcal{N}} } \sum_{i \in {\mathcal{I}}_n } \Big( B^{\mathrm{eng}}_{n, f} x_{i,n,f} \nonumber \\
	& & + \big( \frac{1}{3} ( A^{f}_{\mathrm{tx}}(\bar{\omega}) \beta^{\mathrm{r}}_{\mathrm{tx}} + A^{f}_{\mathrm{rx}}(\bar{\omega})  \beta^{\mathrm{r}}_{\mathrm{rx}} ) \big) y^{\mathrm{r}}_{i,n,f} \nonumber \\	
	& & + ( A^{f}_{\mathrm{km}} \beta^{\mathrm{r}}_{\mathrm{km}} + A^{f}_{\mathrm{si}} \beta^{\mathrm{r}}_{\mathrm{si}} + A^{f}_{\mathrm{md}} \beta^{\mathrm{r}}_{\mathrm{md}} ) z^{\mathrm{r}}_{i,n,f} \nonumber \\
	& & +  e_{(i,n)} ( y^{\mathrm{r}}_{i,n,f} + z^{\mathrm{r}}_{i,n,f} )  \beta^{\mathrm{r}}_{\mathrm{ch}} \Big) \nonumber \\
	& &  + {\mathbb{E}} \left[ {\mathscr{C}}_n ( x_{i,n,f}, y^{\mathrm{r}}_{i, n, f }, z^{\mathrm{r}}_{i,n,f}, \tilde{\omega} ) \right], 	
	\label{eq:sp_obj1} \\
\mbox{where} 
	& & {\mathscr{C}}_n ( y^{\mathrm{r}}_{i, n, f }, z^{\mathrm{r}}_{i,n,f}, \tilde{\omega} ) \nonumber \\ 
	& & =   \min_{x_{i,n,f}, y^{\mathrm{r}}_{i,n,f}, z^{\mathrm{r}}_{i,n,f}}  \sum_{f \in {\mathcal{F}}}  \sum_{n \in {\mathcal{N}} } \sum_{i \in {\mathcal{I}}_n } \nonumber \\
	& & \Big( \big( \frac{1}{3} ( A^{f}_{\mathrm{tx}}(\tilde{\omega})\beta^{\mathrm{e}}_{\mathrm{tx}} + A^{f}_{\mathrm{rx}}(\tilde{\omega}) \beta^{\mathrm{e}}_{\mathrm{rx}} ) \big) y^{\mathrm{e}}_{i,n,f,\tilde{\omega}} \nonumber \\
	& &  + ( A^{f}_{\mathrm{km}}\beta^{\mathrm{e}}_{\mathrm{km}} + A^{f}_{\mathrm{si}}\beta^{\mathrm{e}}_{\mathrm{si}} + A^{f}_{\mathrm{md}}\beta^{\mathrm{e}}_{\mathrm{md}} ) z^{\mathrm{e}}_{i,n,f,\tilde{\omega}} \nonumber \\
	& & + e_{(i,n)} ( y^{\mathrm{e}}_{i,n,f,\tilde{\omega}} + z^{\mathrm{e}}_{i,n,f,\tilde{\omega}} )\beta^{\mathrm{e}}_{\mathrm{ch}}  \nonumber \\
	& & + \big( \frac{1}{3} ( A^{f}_{\mathrm{tx}}(\tilde{\omega}) \beta^{\mathrm{o}}_{\mathrm{tx}} + A^{f}_{\mathrm{rx}}(\tilde{\omega})  \beta^{\mathrm{o}}_{\mathrm{rx}} ) \big) y^{\mathrm{o}}_{i,n,f,\tilde{\omega}} \nonumber \\
	& & + ( A^{f}_{\mathrm{km}} \beta^{\mathrm{o}}_{\mathrm{km}} + A^{f}_{\mathrm{si}} \beta^{\mathrm{o}}_{\mathrm{si}} + A^{f}_{\mathrm{md}} \beta^{\mathrm{o}}_{\mathrm{md}} ) z^{\mathrm{o}}_{i,n,f,\tilde{\omega}} \nonumber \\
	& & + e_{(i,n)} ( y^{\mathrm{o}}_{i,n,f,\tilde{\omega}} + z^{\mathrm{o}}_{i,n,f,\tilde{\omega}} ) \beta^{\mathrm{o}}_{\mathrm{ch}} \Big)  \label{eq:sp_obj2} \\
\mbox{s.t.} 
	%%%%%% routing constraints %%%%%%		
	& & \sum_{ j' \in {\mathcal{O}}_{S_f} }	x_{S_f,j',f}	-	\sum_{ i' \in {\mathcal{I}}_{S_f} }	x_{i',S_f,f} =	1,	f	\in {\mathcal{F}}	\label{eq:sp_rt_const1} \\
	& & \sum_{ i' \in {\mathcal{I}}_{D_f} } x_{i', D_f, f} - \sum_{ j' \in {\mathcal{O}}_{D_f} } x_{D_f, j', f } =	1,	f \in {\mathcal{F}}	 \label{eq:sp_rt_const2} \\
	& & \sum_{ j' \in {\mathcal{O}}_n } x_{n, j', f }	-	\sum_{i' \in {\mathcal{I}}_n } x_{i',n, f}	=	0, \nonumber \\ 
	& &  f \in {\mathcal{F}}, n \in {\mathcal{N}} \setminus \{ S_f, D_f \} \label{eq:sp_rt_const3} \\
	& & \sum_{j' \in {\mathcal{O}}_n } x_{n, j', f } 	\leq	1, n \in {\mathcal{N}}, f \in {\mathcal{F}} \label{eq:sp_rt_const4} \\ 
	& & \sum_{f \in {\mathcal{F}}} \big(   y^{\mathrm{e}}_{i,j, f, \tilde{\omega}}  x_{i,j, f}  \big)	\leq W^{\mathrm{qkd}}_{i,j} (\mathbb{C}), i,j \in {\mathcal{N}}	\label{eq:sp_rt_const5} \\
	& & \sum_{f \in {\mathcal{F}}} (  z^{\mathrm{e}}_{i,j, f, \tilde{\omega}}  x_{i,j, f} )	\leq	W^{\mathrm{kml}}_{i,j} (\mathbb{C}),  i,j \in {\mathcal{N}} \label{eq:sp_rt_const6} \\
	%%%%%% expending and on-demand phases %%%%%%
	& & y^{\mathrm{e}}_{i,n,f,\tilde{\omega}} x_{i,j, f} \leq y^{\mathrm{r}}_{i,n,f} x_{i,j, f}, i,n \in {\mathcal{N}},  f \in {\mathcal{F}} \label{eq:sp_rt_const7} \\
	& & z^{\mathrm{e}}_{i,n,f,\tilde{\omega}} x_{i,j, f} \leq z^{\mathrm{r}}_{i,n,f} x_{i,j, f}, i,n \in {\mathcal{N}},  f  \in {\mathcal{F}} \label{eq:sp_rt_const8} \\ 
	& & ( y^{\mathrm{e}}_{i,n,f,\tilde{\omega}} x_{i,j, f} ) + y^{\mathrm{o}}_{i,n,f,\tilde{\omega}} \geq   P_{f}({\tilde{\omega}}) W^{\mathrm{qkd}}_{f} x_{i,n, f}, \nonumber \\
	& & i,n \in {\mathcal{N}}, f	\in {\mathcal{F}} \label{eq:sp_rt_const9}\\  
	& & (z^{\mathrm{e}}_{i,n,f,\tilde{\omega}} x_{i,j, f} ) + z^{\mathrm{o}}_{i,n,f,\tilde{\omega}} \geq   P_{f}({\tilde{\omega}}) W^{\mathrm{kml}}_{f}  x_{i,n, f}, \nonumber \\
	& & i,n \in {\mathcal{N}}, f	\in {\mathcal{F}} \label{eq:sp_rt_const10} \\
	& & W^{\mathrm{qkd}}(\mathbb{C})_{i,j} \leq W^{\mathrm{qkd}}_{\mathrm{max}}, i,j \in {\mathcal{N}} \label{eq:sp_rt_const11} \\ % the resource of pool must be less than that of available physical bandwidth of network
	& & W^{\mathrm{kml}}(\mathbb{C})_{i,j}  \leq W^{\mathrm{kml}}_{\mathrm{max}}, i,j \in {\mathcal{N}}  \label{eq:sp_rt_const12} \\ % the resource of pool must be less than that of available physical bandwidth of network
	& & x_{i,n,f} \in \{ 0, 1 \}, i,n \in {\mathcal{N}}, f \in {\mathcal{F}} \label{eq:sp_rt_const13}  \\
	& & y^{\mathrm{r}}_{i, n, f }, y^{\mathrm{e}}_{i, n, f, \tilde{\omega} }, y^{\mathrm{o}}_{i, n, f, \tilde{\omega} }, z^{\mathrm{r}}_{i, n, f }, z^{\mathrm{e}}_{i, n, f, \tilde{\omega} }, \nonumber \\
	& & z^{\mathrm{o}}_{i, n, f, \tilde{\omega}} \in \{0,1,2,\dots\}, i,n \in {\mathcal{N}}, f \in {\mathcal{F}} \label{eq:sp_rt_const14}   	
\eeqn

In the proposed system, QKD resources (i.e., QKD and KM wavelengths) are allocated by QKD service providers to SIC applications in the QKD network. In the first stage, QKD resources are allocated in the resource pool of QKD nodes. The actual number of QKD resources in the resource pool is determined in the second stage. The SP model considers the uncertainties (i.e., the number of \emph{scenarios}), including the number of QKD devices required by the SIC nodes. In the QKD network, the secret-key rates are uncertain. Let $\tilde{\kappa}_{f}$ denote a random variable representing the secret key rate required for the QKD-SIC chain request $f$. Let $\varphi$ denote the size of the scenario spaces $\mathcal{K}_{f}$.
\beqn
	 & & \tilde{\omega} \in \big\{  \tilde{\kappa}_{f} \big| \tilde{\kappa}_{f} \in \mathcal{K}_{f} \big\}, \nonumber  \\
     & & \mathcal{K}_{f} = \{ \tilde{\kappa}_{f_1}, \tilde{\kappa}_{f_2}, \tilde{\kappa}_{f_3},\dots,\tilde{\kappa}_{f_{\varphi}} \}. \label{eq:omega-random-variable}	
\eeqn

The objective of the SP model for QKD resource allocation under uncertainty expressed in (\ref{eq:sp_obj1}) is to minimize the total cost of deploying QKD resources, including MDI-QTxs, MDI-QRxs, LKMs, SIs, MUX /DEMUX components, and QKD and KM links.

The constraint (\ref{eq:sp_rt_const1}) ensures that the number of outgoing routes is larger than the number of incoming routes when the node is the source node $S_f$ of the QKD-SIC chain request $f$. The constraint (\ref{eq:sp_rt_const2}) ensures that the number of incoming routes is larger than the number of outgoing routes if the node is the destination node $D_f$ of the QKD-SIC chain request $f$. The constraint (\ref{eq:sp_rt_const3}) ensures that the number of outgoing routes must equal the number of incoming routes if the node is the intermediate node of the QKD-SIC chain request $f$. The constraint (\ref{eq:sp_rt_const4}) ensures that there is no loop for any QKD-SIC chain request, meaning that there is only one outgoing route for the QKD-SIC chain request of any node. The constraint (\ref{eq:sp_rt_const5}) ensures that the QKD wavelengths of all QKD-SIC chain requests of any node must not exceed the available QKD wavelengths in the QKD resource pool (i.e., $W^{\mathrm{qkd}}_{i,j} (\mathbb{C})$). The constraint (\ref{eq:sp_rt_const6}) ensures that the KM wavelengths of all QKD-SIC chain requests from any node must not exceed the available KM wavelengths in the QKD resource pool (i.e., $W^{\mathrm{kml}}_{i,j} (\mathbb{C})$).

The constraint (\ref{eq:sp_rt_const7}) ensures that the number of QKD wavelengths utilized is less than or equal to the number of QKD wavelengths reserved. The constraint (\ref{eq:sp_rt_const8}) ensures that the number of KM wavelengths utilized is less than or equal to the number of KM wavelengths reserved. The constraint (\ref{eq:sp_rt_const9}) ensures that the number of QKD wavelengths used plus the number of on-demand QKD wavelengths must satisfy the security requirements (i.e., secret-key rates). The constraint (\ref{eq:sp_rt_const10}) ensures that the number of KM wavelengths used plus the number of on-demand KM wavelengths must satisfy the requirements. The constraints (\ref{eq:sp_rt_const11}) and (\ref{eq:sp_rt_const12}) ensure that the QKD and KM wavelength resource pools do not exceed the maximum available QKD and wavelength resource pools, respectively. The constraints (\ref{eq:sp_rt_const13}) and (\ref{eq:sp_rt_const14}) are the binary and integer variables, respectively.

\subsection{Deterministic Equivalent Formulation}
\label{subsec:deterministic-evuivalent}

%%%%%%%%%%%%%%% deterministic equivalent formulation version %%%%%%%%%%%%%%%%%%%%
\beqn
	v (\mathbb{C}) & & = \min_{x_{i,n,f}, y^{\mathrm{r}}_{i,n,f}, z^{\mathrm{r}}_{i,n,f}}  \sum_{f \in {\mathcal{F}}} \sum_{n \in {\mathcal{N}} } \sum_{i \in {\mathcal{I}}_n } \Big( B^{\mathrm{eng}}_{n, f} x_{i,n,f} \nonumber \\
	& & + \big( \frac{1}{3} ( A^{f}_{\mathrm{tx}}(\bar{\omega}) \beta^{\mathrm{r}}_{\mathrm{tx}} + A^{f}_{\mathrm{rx}}(\bar{\omega})  \beta^{\mathrm{r}}_{\mathrm{rx}} ) \big) y^{\mathrm{r}}_{i,n,f} \nonumber \\
	& & + ( A^{f}_{\mathrm{km}} \beta^{\mathrm{r}}_{\mathrm{km}} + A^{f}_{\mathrm{si}} \beta^{\mathrm{r}}_{\mathrm{si}} + A^{f}_{\mathrm{md}} \beta^{\mathrm{r}}_{\mathrm{md}} ) z^{\mathrm{r}}_{i,n,f} \nonumber \\
	& & + e_{(i,n)} ( y^{\mathrm{r}}_{i,n,f} + z^{\mathrm{r}}_{i,n,f} ) \beta^{\mathrm{r}}_{\mathrm{ch}} \Big) \nonumber \label{eq:sp_obj} \\
	& & + \sum_{f \in {\mathcal{F}}} \mathbb{P}_{f}(\omega) \sum_{n \in {\mathcal{N}} } \sum_{i \in {\mathcal{I}}_n } \nonumber \\
	& & \Big(  \big( \frac{1}{3} ( A^{f}_{\mathrm{tx}}(\omega)\beta^{\mathrm{e}}_{\mathrm{tx}} + A^{f}_{\mathrm{rx}}(\omega) \beta^{\mathrm{e}}_{\mathrm{rx}} ) \big) y^{\mathrm{e}}_{i,n,f,\omega}  \nonumber \\
	& & + ( A^{f}_{\mathrm{km}}\beta^{\mathrm{e}}_{\mathrm{km}} + A^{f}_{\mathrm{si}}\beta^{\mathrm{e}}_{\mathrm{si}} + A^{f}_{\mathrm{md}}\beta^{\mathrm{e}}_{\mathrm{md}} ) z^{\mathrm{e}}_{i,n,f,\omega} \nonumber \\ 
	& & + e_{(i,n)} ( y^{\mathrm{e}}_{i,n,f,\omega} + z^{\mathrm{e}}_{i,n,f,\omega} )\beta^{\mathrm{e}}_{\mathrm{ch}} \nonumber \\
	& & + \big( \frac{1}{3} ( A^{f}_{\mathrm{tx}}(\omega) \beta^{\mathrm{o}}_{\mathrm{tx}} + A^{f}_{\mathrm{rx}}(\omega)  \beta^{\mathrm{o}}_{\mathrm{rx}} ) \big) y^{\mathrm{o}}_{i,n,f,\omega} \nonumber \\ 
	& & +  ( A^{f}_{\mathrm{km}} \beta^{\mathrm{o}}_{\mathrm{km}} + A^{f}_{\mathrm{si}} \beta^{\mathrm{o}}_{\mathrm{si}} + A^{f}_{\mathrm{md}} \beta^{\mathrm{o}}_{\mathrm{md}} ) z^{\mathrm{o}}_{i,n,f,\omega} \nonumber \\
	& & + e_{(i,n)} ( y^{\mathrm{o}}_{i,n,f,\omega} + z^{\mathrm{o}}_{i,n,f,\omega} ) \beta^{\mathrm{o}}_{\mathrm{ch}}  \Big)  \label{def:sp_obj} \\
\mbox{s.t.} 
	%%%%%% routing constraints %%%%%%		
	& & \sum_{ j' \in {\mathcal{O}}_{S_f} }	x_{S_f,j',f}	-	\sum_{ i' \in {\mathcal{I}}_{S_f} }	x_{i',S_f,f} =	1,	 f	\in {\mathcal{F}}	\label{eq:def_rt_const1} \\
	& & \sum_{ i' \in {\mathcal{I}}_{D_f} } x_{i', D_f, f} - \sum_{ j' \in {\mathcal{O}}_{D_f} } x_{D_f, j', f } =	1,	 f \in {\mathcal{F}}	 \label{eq:def_rt_const2} \\
	& & \sum_{ j' \in {\mathcal{O}}_n } x_{n, j', f }	-	\sum_{i' \in {\mathcal{I}}_n } x_{i',n, f}	=	0, \nonumber \\
	& & f	\in {\mathcal{F}}, n \in {\mathcal{N}} \setminus \{ S_f, D_f \} \label{eq:def_rt_const3} \\
	& & \sum_{j' \in {\mathcal{O}}_n } x_{n, j', f } 	\leq	1,	 n \in {\mathcal{N}}, f \in {\mathcal{F}} \label{eq:def_rt_const4} \\ 
	& & \sum_{f \in {\mathcal{F}}} \big(  y^{\mathrm{e}}_{i,j, f, \omega}  x_{i,j, f}  \big) \leq W^{\mathrm{qkd}}_{i,j} (\mathbb{C}), \nonumber \\
	& & i,j \in {\mathcal{N}}, \forall \omega \in \Omega_{f} \label{eq:def_rt_const5} \\
	& & \sum_{f \in {\mathcal{F}}} ( z^{\mathrm{e}}_{i,j, f, \omega}  x_{i,j, f} ) \leq	W^{\mathrm{kml}}_{i,j} (\mathbb{C}), \nonumber \\
	& & i,j \in {\mathcal{N}}, \forall \omega \in \Omega_{f}	\label{eq:def_rt_const6} \\
	%%%%%% expending and on-demand phases %%%%%%
	& & y^{\mathrm{e}}_{i,n,f,\omega} x_{i,j, f} \leq y^{\mathrm{r}}_{i,n,f} x_{i,j, f}, \nonumber \\
	& & i,n \in {\mathcal{N}},  f \in {\mathcal{F}}, \forall \omega \in \Omega_{f} \label{eq:def_rt_const7} \\
	& & z^{\mathrm{e}}_{i,n,f,\omega} x_{i,j, f} \leq z^{\mathrm{r}}_{i,n,f} x_{i,j, f}, \nonumber \\
	& & i,n \in {\mathcal{N}},  f  \in {\mathcal{F}}, \forall \omega \in \Omega_{f} \label{eq:def_rt_const8}\\ 
	& & ( y^{\mathrm{e}}_{i,n,f,\omega} x_{i,j, f}) + y^{\mathrm{o}}_{i,n,f,\omega}  \geq \nonumber \\
	& & \sum_{\omega \in \Omega_{f}} ( P_{f}({\omega}) W^{\mathrm{qkd}}_{f} x_{i, n, f} ), i,n \in {\mathcal{N}}, f \in {\mathcal{F}} \label{eq:def_rt_const9} \\  
	& & (z^{\mathrm{e}}_{i,n,f,\omega} x_{i,j, f})  + z^{\mathrm{o}}_{i,n,f,\omega}  \geq \nonumber \\
	& & \sum_{\omega \in \Omega_{f}} ( P_{f}({\omega}) W^{\mathrm{kml}}_{f}  x_{i,n, f} ), i,n \in {\mathcal{N}}, f	\in {\mathcal{F}} \label{eq:def_rt_const10} \\
	& & W^{\mathrm{qkd}}_{i,j} (\mathbb{C}) \leq W^{\mathrm{qkd}}_{\mathrm{max}}, i,j \in {\mathcal{N}} \label{eq:def_rt_const11} \\ % the resource of pool must be less than that of available physical bandwidth of network
	& & W^{\mathrm{kml}}_{i,j}(\mathbb{C}) \leq W^{\mathrm{kml}}_{\mathrm{max}}, i,j \in {\mathcal{N}} \label{eq:def_rt_const12} \\ % the resource of pool must be less than that of available physical bandwidth of network
	& & x_{i,n,f} \in \{ 0, 1 \}, i,n \in {\mathcal{N}}, f \in {\mathcal{F}}  \label{eq:def_rt_const13} \\
	& & y^{\mathrm{r}}_{i, n, f }, y^{\mathrm{e}}_{i, n, f, \omega }, y^{\mathrm{o}}_{i, n, f, \omega }, z^{\mathrm{r}}_{i, n, f }, z^{\mathrm{e}}_{i, n, f, \omega }, z^{\mathrm{o}}_{i, n, f, \omega} \in \nonumber \\
	& & \{0,1,2,\dots\}, i,n \in {\mathcal{N}}, f \in {\mathcal{F}}, \forall \omega \in \Omega_{f}  \label{eq:def_rt_const14}  	
\eeqn

The SP with the random variable $\tilde{\omega}$ in (\ref{eq:sp_obj1}) - (\ref{eq:sp_rt_const14}) can be transformed into the deterministic equivalence problem \cite{Brige1997} as expressed in (\ref{eq:sp_obj}) - (\ref{eq:def_rt_const14}). The random variable $\tilde{\omega}$ can be represented by a scenario. The scenario is a realization of a random variable, which is denoted by $\omega$. The value of the random variable can be taken from a set of scenarios. Let $\Psi$ and $\Omega_{f}$ as defined in (\ref{eq:scenario_space1}) and (\ref{eq:scenario_space2}), respectively, be the set of all scenarios of each requirement (i.e., a scenario space) and the set of all scenarios of request $f$, respectively. Let $K$ denote the maximum required secret-key rate of the request $f$.
\beqn
 &&\Psi = {\displaystyle \prod_{f \in \mathcal{F}}} \Omega_{f} = \Omega_{1} \times \Omega_{2} \times \cdots \times \Omega_{|\mathcal{F}|} 	\label{eq:scenario_space1}	\\
 & & \Omega_f  = \{ 0, 1, 2,\dots, K \} 	\label{eq:scenario_space2}
\eeqn

The expectation $\mathbb{E}[\cdot]$ of the SP in (\ref{eq:sp_obj1}) can be represented by the weighted sum of scenarios and their probabilities $\mathbb{P}_{f}(\omega)$.

The objective function (\ref{eq:sp_obj}) is to minimize the deployment cost. The decision variables $y^{\mathrm{e}}_{i, n, f, \omega }$, $y^{\mathrm{o}}_{i, n, f, \omega }$, $z^{\mathrm{e}}_{i, n, f, \omega }$, and $z^{\mathrm{o}}_{i, n, f, \omega}$ are under $\omega$ (i.e., $\omega \in \Omega$) which implies that the values of demands are available when $\omega$ is observed. The constraints (\ref{eq:def_rt_const1}), (\ref{eq:def_rt_const2}), (\ref{eq:def_rt_const3}), and (\ref{eq:def_rt_const4}) are the routing constraints, which have the same meanings as (\ref{eq:sp_rt_const1}), (\ref{eq:sp_rt_const2}), (\ref{eq:sp_rt_const3}), and (\ref{eq:sp_rt_const4}), respectively. The constraints (\ref{eq:def_rt_const5}) and (\ref{eq:def_rt_const6}) guarantee that the utilization of QKD and KM wavelengths does not exceed the QKD and KM wavelength capacity, which are similar to (\ref{eq:sp_rt_const5}) and (\ref{eq:sp_rt_const6}). The constraints (\ref{eq:def_rt_const7}) and  (\ref{eq:def_rt_const8}) guarantee that the expended QKD and KM wavelengths respectively must not exceed the reserved QKD and KM wavelengths, which have the same meanings as (\ref{eq:sp_rt_const7}) and (\ref{eq:sp_rt_const8}). The constraints (\ref{eq:def_rt_const9}) and  (\ref{eq:def_rt_const10}) guarantee that all demands are satisfied, which are similar to (\ref{eq:sp_rt_const9}) and  (\ref{eq:sp_rt_const10}), respectively. The constraints (\ref{eq:def_rt_const11}) and (\ref{eq:def_rt_const12}) are the capacity constraints of QKD and KM wavelength resource pools which have the same meanings as (\ref{eq:sp_rt_const11}) and  (\ref{eq:sp_rt_const12}). The constraints in (\ref{eq:def_rt_const13}) and  (\ref{eq:def_rt_const14}) indicate that all decision variables are non-negative integers.

\section{Cost Management Among QKD Service Providers}
\label{sec:cost-management}

 When the QKD resource allocation for maritime transportation applications is performed, the cooperative QKD service providers in coalition $\mathbb{C}$ share the deployment cost created from satisfying the semantic requirements. In this section, we present the cost management to apply the concept of Shapley value to determine the deployment cost-sharing that each of cooperative QKD service providers will obtain. 
 
\subsection{Shapley Value}
\label{subsec:shapley}

The concept of Shapley value~\cite{peleg2007introduction} is applied to determine the deployment cost-sharing among cooperative QKD service providers\footnote{In the remainder of this paper, ``QKD service provider'' and ``provider'' are used interchangeably.}. The cooperative providers form a coalition to establish the shared QKD resource pool. Given the characteristic function $v(\cdot)$ in (\ref{eq:sp_obj1}) and (\ref{def:sp_obj}), the Shapley value of QKD service provider $s$ is expressed in as follows:
\beq
	\vartheta_{s}(v) = {\displaystyle \sum_{\mathbb{D} \subseteq {\mathbb{C}} \setminus \{s\}}} 
	\frac{ \mathbb{|D|}! ( |{\mathbb{C}}| - \mathbb{|D|} - 1)!} { |{\mathbb{C}}|!} \big( v(\mathbb{D} \cup \{s\}) - v(\mathbb{D}) \big ). \label{Shapley_eq}
\eeq 
The cost management of the cooperative QKD service providers in coalition ${\mathbb{C}}$ based on the Shapley value can achieve four desirable properties as follows:
\begin{enumerate}
    \item \emph{Efficiency:} It is $\displaystyle \sum_{ s \in {\mathbb{C}}} \vartheta_{s}(v) = v({\mathbb{C}})$ which means that the sum of deployment cost of all cooperative QKD service providers is minimum. 
    \item \emph{Symmetry:} $\vartheta_{s}(v)$ is equal to $\vartheta_{l}(v)$ if $v(\mathbb{D} \cup \{s\}) = v(\mathbb{D} \cup \{l\})$ contains all QKD service providers $s$ and $l$ given all coalitions $\mathbb{D}$ of other cooperative QKD service providers. This means that the cost shared between QKD service providers $s$ and $l$ is equal if they make an identical contribution to the coalition.
    \item \emph{Dummy:} $\vartheta_{s}(v)$ is 0 if $v(\mathbb{D}) = v(\mathbb{D} \cup \{s\})$ contains all coalitions $\mathbb{D}$ of all cooperative QKD service providers except provider $s$. This means that the cost shared by QKD service provider $s$ is zero if QKD service provider $s$ does not make a contribution to the total cost of the coalition.
    \item \emph{Additivity:} $\vartheta(u+v) = \vartheta(v+u) = \vartheta(u) + \vartheta(v)$ if $u$ and $v$ are the characteristic functions.
\end{enumerate}

With the Shapley value, individual fairness is achieved~\cite{peleg2007introduction}. In particular, the cost shared by the cooperative QKD service provider is not more than the cost of the QKD service provider who does not cooperate with other QKD service providers (i.e., $\vartheta_{s}(v) \leq v(\{s\})$). In addition, the uniqueness of the Shapley value is ensured~\cite{peleg2007introduction}.

\section{Cooperation Formation Among QKD Service Providers}
\label{sec:cooperation-formation}

In this section, we investigate the condition that QKD service providers aim to minimize their own costs. Therefore, QKD service providers may form a coalition and create a resource pool (i.e., QKD and KM wavelengths). We analyze the cooperation formation of QKD service providers by considering their individual costs. 
%%%%%%%%%%%%%%%%%%%%%%%%%%%%%%%%%%%%%%%%%%%%%%%%%%%%%%%%%%%%%%%%%%%%%%%%%%%%%%%%%%%%%%
\subsection{Cooperation Model}
\label{subsec:cooperation-model}

We detail the model of cooperation formation of the QKD service providers with the following information. Let a set of all providers (i.e., players) be denoted by ${\mathbb{G}}$. There is an agreement among providers in ${\mathbb{C}}$ on cooperation (i.e., ${\mathbb{C}} \subseteq {\mathbb{G}}$). ${\mathbb{C}}$ is a set of cooperative providers such that the characteristic function of cooperative providers is $v({\mathbb{C}})$, which is given by (\ref{eq:sp_obj}). For example, given three providers, a set of providers is ${\mathbb{G}} = \{ 1, 2, 3\}$. The cooperation can be defined as ${\mathbb{C}}$ $\in \{ \{ \{1\}, \{2\}, \{3\} \}, \{\{1,2\}, \{3\} \}, \{\{1,3\}, \{2\}\}, \{\{2,3\}, \{1\}\},$ $\{\{1,2,3\}\} \}$. The coalition game with a set of players that are providers can be defined as $({\mathbb{G}},\delta_s(\cdot))$. Therefore, we can formulate a coalition game to model and obtain the equilibrium strategies. The strategy of each provider is to cooperate with other providers. Let $p_{s,l}$ denote the cooperation. If $p_{s,l} = 1$ which means that providers $s$ and $l$ cooperate, and $p_{s,l}=0$ otherwise. Let $\mathbf{p}_{s} \in \mathcal{P}_{s}$ denote the strategy of provider $s$. Let $\mathbf{p}^{*}_{s} \in \mathcal{P}_{s}$ denote the cooperation equilibrium strategy. Let $\mathbf{p}^{*}_{-s} \in \displaystyle{\prod_{l \in \mathbb{G} \setminus \{s\} } \mathcal{P}_{l}}$ denote the cooperation equilibrium strategies of all providers except the strategy of provider $s$. Therefore, the strategy space of provider $s$ cooperating with other providers is expressed in (\ref{eq:strategyspace}). The cooperation structure ${\mathbb{C}}$ is expressed in (\ref{eq:mappedgraph}). Therefore, the cooperation equilibrium of the cooperation formation of providers is expressed in (\ref{eq_nash}). 
\beqn
 	& & \mathcal{P}_{s} = \Big\{( p_{s,1},\dots, p_{s,l-1}, p_{s,l}, p_{s,l+1},\dots, p_{s,|{\mathbb{G}} |})  \nonumber \\
    & & | p_{s,l} \in \{0,1\}, l \in {\mathbb{G}} \setminus \{s\} \Big\} \label{eq:strategyspace} \\
    & & p_{s,l} =
 	\left\{ \begin{array}{rcl}
			 1 ~\mathrm{if} ~ s \in {\mathbb{C}} ~~ \mathrm{AND} ~~ l \in {\mathbb{C}}  \\
 		 0 ~\mathrm{if} ~ s \notin {\mathbb{C}} ~~ \mathrm{OR} ~~ l \notin {\mathbb{C}} \;\;\; \\
			\end{array}\right.  \label{eq:mappedgraph}  \\
	& & \delta_{s} (\mathbf{p}^{*}_{s}, \mathbf{p}^{*}_{-s}) \leq \delta_{s}(\mathbf{p}_{s}, \mathbf{p}^{*}_{-s}),\quad \forall s  \label{eq_nash}
\eeqn
$\delta_{s} (\cdot)$ is the cost of provider $s$ received from the QKD resource allocation and cost management, i.e., $\delta_{s} (\cdot) = \vartheta_{s}$ in~(\ref{Shapley_eq}).

\subsection{Dynamics of Coalition Formation}
\label{subsec:dynamics-of-coalition-formation}

The cooperation formation equilibrium among providers is determined based on best-response dynamics. In particular, the provider iteratively makes a decision to form a cooperation. The decision iteration is denoted by $\sigma$ (i.e., $\sigma = 1, 2, 3, \dots, N $) where $N$ is a maximum number of iterations. The strategy of provider $s$ in the decision iteration $\sigma$ is denoted by $\mathbf{p}_{s}(\sigma)$. The strategies of all providers except the strategy of provider $s$ in decision iteration $\sigma - 1$ are denoted by $\mathbf{p}_{-s}(\sigma - 1)$. In each decision iteration, the provider assesses the new strategy and then changes to the new strategy with the lowest cost. The strategy of provider $s$ in decision iteration $\sigma$ (i.e., $\mathbf{p}_{s}(\sigma)$) can be determined as follows:
\beqn
	\mathbf{p}_{s}(\sigma) \in \mathrm{arg} \; \min_{\mathbf{p}_{s} \in \mathcal{P}_{s}} \;\; \delta_{s}(\mathbf{p}_{s}, \mathbf{p}_{-s}(\sigma - 1)). \label{eq_strategy}
\eeqn
In~(\ref{eq_strategy}), given the knowledge of strategies of other providers in the prior decision iteration (i.e., $ \mathbf{p}_{-s}(\sigma - 1)$), the best new strategy is selected by the provider $s$. However, the provider may make a mistake with a small probability (i.e., $\aleph$). 

From~(\ref{eq_strategy}), discrete-time Markov chain can be applied to model the strategy adaptation of the cooperation formation~\cite{Goyal2005}. Under all cooperations of all providers, the finite state space of the Markov chain is expressed as $ \Theta = \displaystyle{\prod_{s \in \mathbb{G}} \mathcal{P}_{s} = \mathcal{P}_{1}\times \dots \times \mathcal{P}_{|\mathbb{G}|}}$. It is assumed that it has symmetric cooperation which is $p_{s,l} = p_{l,s}$. The strategy $\mathbf{p}_{s}(\cdot)$ in~(\ref{eq_strategy}) contains the cooperations of provider $s$ and is part of the state $\tau$, i.e., $\tau \in \Theta$.  For the transition probability of Markov chain, $\tau$ and $\tau^{'}$ are the current state and the next state, respectively. Let $\tau = ( p_{s,1},\dots, p_{s,l}, \dots, p_{s,|\mathbb{G}|}), \tau \in \Theta $ denote the current state. Let $\tau^{'} = ( p^{'}_{s,1},\dots, p^{'}_{s,l},\dots, p^{'}_{s,|\mathbb{G}|}), \tau^{'} \in \Theta$ denote the next state. The set of providers associated with the state change from $\tau$ to $\tau^{'}$  is defined in (\ref{set_of_orgs}).
\beqn
	\mathcal{Z}_{\tau, \tau^{'}} = \{ s | p_{s,l} \neq p^{'}_{s,l}, s \neq l, s,l \in \mathbb{G} \}  \label{set_of_orgs}  
\eeqn
The transition probability from $\tau$ to $\tau^{'}$ is expressed in (\ref{transition_prob}). 
\beqn
	\mathbf{T}_{\tau, \tau^{'}} = \lambda^{|\mathcal{Z}_{\tau, \tau^{'}}|} (1 - \lambda)^{{|\mathbb{G}|} - {|\mathcal{Z}_{\tau, \tau^{'}}|}} \prod_{s \in \mathcal{Z}_{\tau, \tau^{'}}} \Xi_{s}(\tau, \tau^{'}) , \label{transition_prob} 
\eeqn
where $\lambda$ is a probability of the provider updating the strategy in an iteration of the decision. The probability of changing the strategy of provider $s$ in an iteration of the decision is expressed in (\ref{best-response_rule}). Let $\delta_{s}(\tau)$ denote the cost that is the function of the strategies of all providers. In (\ref{best-response_rule}), the provider can change to the strategy that produces lower cost (i.e., $\delta_{s}(\tau^{'}) < \delta_{s}(\tau)$). However, the provider may irrationally change to the new strategy with probability $\aleph$.
\beqn
	\Xi_{s}(\tau, \tau^{'}) =
			\left\{ \begin{array}{rcl}
			 1 - \aleph ~\mathrm{if} ~\delta_{s}(\tau^{'}) < \delta_{s}(\tau) \\
 			 \aleph ~\mathrm{otherwise} \quad\quad\quad\;
			\end{array}\right. \label{best-response_rule}
\eeqn 
%%%%%%%%%%%%%%%%%%%%%%%%%%%%%%%%%%%%%%%%%%%%%%%%%%%%%%%%%%%%%%%%%%%%%%%%%%%%%%%%%%%%%%%%%%%%%%
 
\section{Performance Evaluation}
\label{sec:performance-evaluation}

In this section, we perform experiments with the following considerations. First, we consider that the proposed SP model can obtain the efficient routing to satisfy the QKD-SIC chain requests\footnote{In the remainder of this paper, ``QKD-SIC chain request'' and ``QKD chain request'' are used interchangeably.}. Each routing consists of QKD and KM wavelengths in the reservation and the on-demand phases of the SP model. The routing results are shown in Section \ref{subsubsec:routing}. Second, we investigate the secret-key rates which have the effect on increasing the deployment cost and consider that the deployment cost can be decreased by optimally reserving the number of QKD and KM wavelengths in advance. The results are presented in Section \ref{subsubsec:cost-stucture-analysis}. 

Third, we show the upper and lower bounds of the SP model with the consideration that if the gap between the solutions of the bound and the solution of the SP model is small enough, we can apply the upper and lower bounds to obtain the satisfied solutions instead of the SP model. This is due to the fact that formulations of the upper and lower bounds are much lower complexity than the formulation of SP model since the numbers of variables and constraints of formulations of the upper and lower bounds are less than that of formulation of SP model. In addition, we compare the SP model with the CO-QBN algorithm \cite{y-cao-hybrid-trusted2021} with different quantitative analyzes to show the performance of the SP model. The results are shown in Section \ref{subsubsec:performance-ev-various-parameters}. Fourth, we show the SP model with a cooperative game and examine the impact of shared QKD and KM wavelengths. In this experiment, we consider that, with a large number of secret-key rates, the costs of cooperative providers are significantly lower than that of noncooperative providers. The results are shown in Section \ref{subsubsec:impact-of-available-qkd-km-wave}.

Finally, we investigate and show how the coalition structure affects the costs of QKD service providers. The stable coalition structure, which is a solution to the cooperative game (i.e., the cooperation equilibrium), can be determined by the Shapley value, a fair cost-sharing concept. In addition, we investigate how the cost of shared QKD and KM wavelengths and cooperation affect the stable cooperative structure. The reason for this is that we want to know where the cooperation structure is stable when the cost increases. The results are shown in Section \ref{subsubsec:cooperation-formulation-cost}.

\subsection{Parameter Setting}
\label{subsec:parameter-setting}

We consider the SIC over QKD networks as shown in Fig.\ref{fig:system-model}. Specifically, there are three QKD service providers. Providers 1, 2, and 3 have 10, 15, and 20 available QKD wavelengths, respectively. Providers 1, 2, and 3 have 40, 55, and 65 available KM wavelengths, respectively. We perform experiments using the NSFNET and USNET topologies \cite{y-cao-hybrid-trusted2021}. The distance $D$ between two QTxs is set to 160 km \cite{y-cao-hybrid-trusted2021}. We set $W^{\mathrm{qkd}}_{\mathrm{max}} = 1,000$ and $W^{\mathrm{kml}}_{\mathrm{max}} = 300$ for the maximum number of wavelengths for a QKD link between node $i$ and node $j$ and the maximum number of wavelengths for a KM link between node $i$ and node $j$, respectively. We consider the Shapley value for cost-sharing between providers. We implement and solve the SP model using the GAMS/CPLEX solver \cite{Gams}. 

%The amount of energy required for transferring traffic of request $f$ going through node $n$ to be = 1, i.e., $B^{\mathrm{eng}}_{n, f} = 1$.

For the SP model, we consider the random number of secret-key rates with uniform distribution for ease of presentation. We consider the cost values of five QKD network components which are composed of MDI-QTXs, MDI-QRXs, LKMs, SIs, MUX/DEMUX components, and QKD and KM links. For the reservation phase, we apply these cost values from \cite{y-cao-hybrid-trusted2021}. The cost values of the components can be presented in Table \ref{table:parameter-setting}.

%%%%%%%%%%%%%%%%%%%%%%%%%%% Table %%%%%%%%%%%%%%%%%%%%%%%%%%%%%%%%%%%%%%%%%%%%%%%%%%%
\begin{table}[htb] \footnotesize \caption{Reservation, utilization, and on-demand cost values}
\label{table:parameter-setting}
\centering
\scalebox{0.9}{\begin{tabular}{|l|l|l|l|l|l|}\hline
{\bf Notations} & {\bf Values(\$)} & {\bf Notations} & {\bf Values(\$)} & {\bf Notations} & {\bf Values(\$)}	\\ \hline
  $\beta^{\mathrm{r}}_{\mathrm{tx}}, \beta^{\mathrm{e}}_{\mathrm{tx}}$ & 1,500 & $\beta^{\mathrm{r}}_{\mathrm{rx}}, \beta^{\mathrm{e}}_{\mathrm{rx}}$ & 2,250 & $\beta^{\mathrm{r}}_{\mathrm{km}}, \beta^{\mathrm{e}}_{\mathrm{km}}$ & 1,200 \\ \hline
  $\beta^{\mathrm{r}}_{\mathrm{si}}, \beta^{\mathrm{e}}_{\mathrm{si}}$ & 150   & $\beta^{\mathrm{r}}_{\mathrm{md}}, \beta^{\mathrm{e}}_{\mathrm{md}}$ & 300   &  $\beta^{\mathrm{r}}_{\mathrm{ch}}, \beta^{\mathrm{e}}_{\mathrm{ch}}$ & 1   \\ \hline
  $\beta^{\mathrm{o}}_{\mathrm{tx}}$ & 6,000  & $\beta^{\mathrm{o}}_{\mathrm{rx}}$  & 9,000 & $\beta^{\mathrm{o}}_{\mathrm{km}}$ & 3,000   \\ \hline
  $\beta^{\mathrm{o}}_{\mathrm{si}}$ & 500    & $\beta^{\mathrm{o}}_{\mathrm{md}}$  & 900   & $\beta^{\mathrm{o}}_{\mathrm{ch}}$ & 4    \\ \hline
%  $W^{\mathrm{qkd}}_{i,j}$ & 150  & $W^{\mathrm{kml}}_{i,j}$ & 50 \\ \hline
\end{tabular}}
\end{table}

\subsection{Numerical Results}
\label{subsubsec:numerical-results}

\subsubsection{Routing}
\label{subsubsec:routing}

%%%%%%%%%%%%%%%%%%%%%%%%%%%%%%%%%%%%%%%%%%%%%%%%%%%%%%%%%%%%%%%%%%%%%%%%%%%%%%%%%%
%routing
\begin{figure}[htb]
%\begin{center}
\centering
\captionsetup{justification=centering}
$\begin{array}{c} \epsfxsize=3.3 in \epsffile{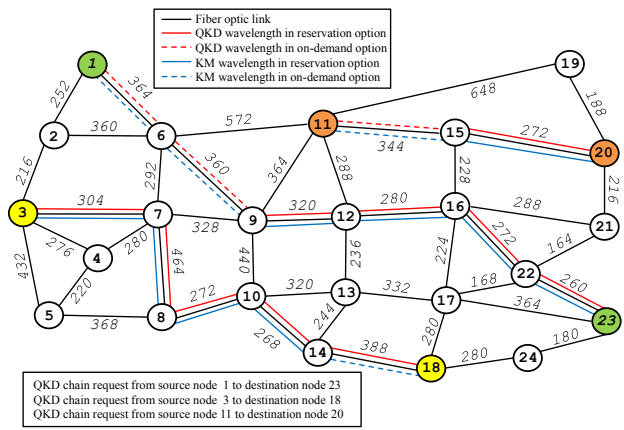} \\
\end{array}$
\caption{An example of three QKD-SIC chain requests in USNET topology.} 
\label{fig:three-routings-fiber}
%\end{center}
\end{figure}

%%%%%%%%%%%%%%%%%%%%%%%%%%%%%%%%%%%%%%%%%%%%%%%%%%%%%%%%%%%%%%%%%%%%%%%%%%%%%%%%%%

Figure \ref{fig:three-routings-fiber} illustrates the solutions of the SP model that satisfy three QKD-SIC chain requests. In the solutions, the SP model can allocate QKD and KM wavelengths (i.e., resources) in reservation and on-demand phases for the requests in fiber optic networks. In Fig. \ref{fig:three-routings-fiber}, each request (i.e., a route) utilizes QKD and KM wavelength within a fiber optic link. To obtain the optimal deployment cost, it is interesting to mention that both QKD and KM wavelengths in reservation and on-demand phases are utilized along the routes. For example, with the QKD-SIC chain request from source node 1 to destination node 23, the optimal deployment cost can be obtained from the route consisted of nodes  $1 \rightarrow 6 \rightarrow 9 \rightarrow 12 \rightarrow 16 \rightarrow 22 \rightarrow 23$. In particular, QKD and KM wavelengths in the on-demand phase are utilized from node 1 to node 6 (i.e., $1 \rightarrow 6$) and from node 6 to node 9 (i.e., $6 \rightarrow 9$) while the rest of the route utilize the QKD and KM wavelengths in the reservation phase.  

\subsubsection{Cost Structure Analysis}
\label{subsubsec:cost-stucture-analysis}

%%%%%%%%%%%%%%%%%%%%%%%%%%%%%%%%%%%%%%%%%%%%%%%%%%%%%%%%%%%%%%%%%%%%%%%%%%%%%%%%%%
\begin{figure*}[htb]
 \centering
 \captionsetup{justification=centering}
 \subfloat[The deployment cost.]{\label{fig:usnet-cost}\includegraphics[width=0.25\textwidth]{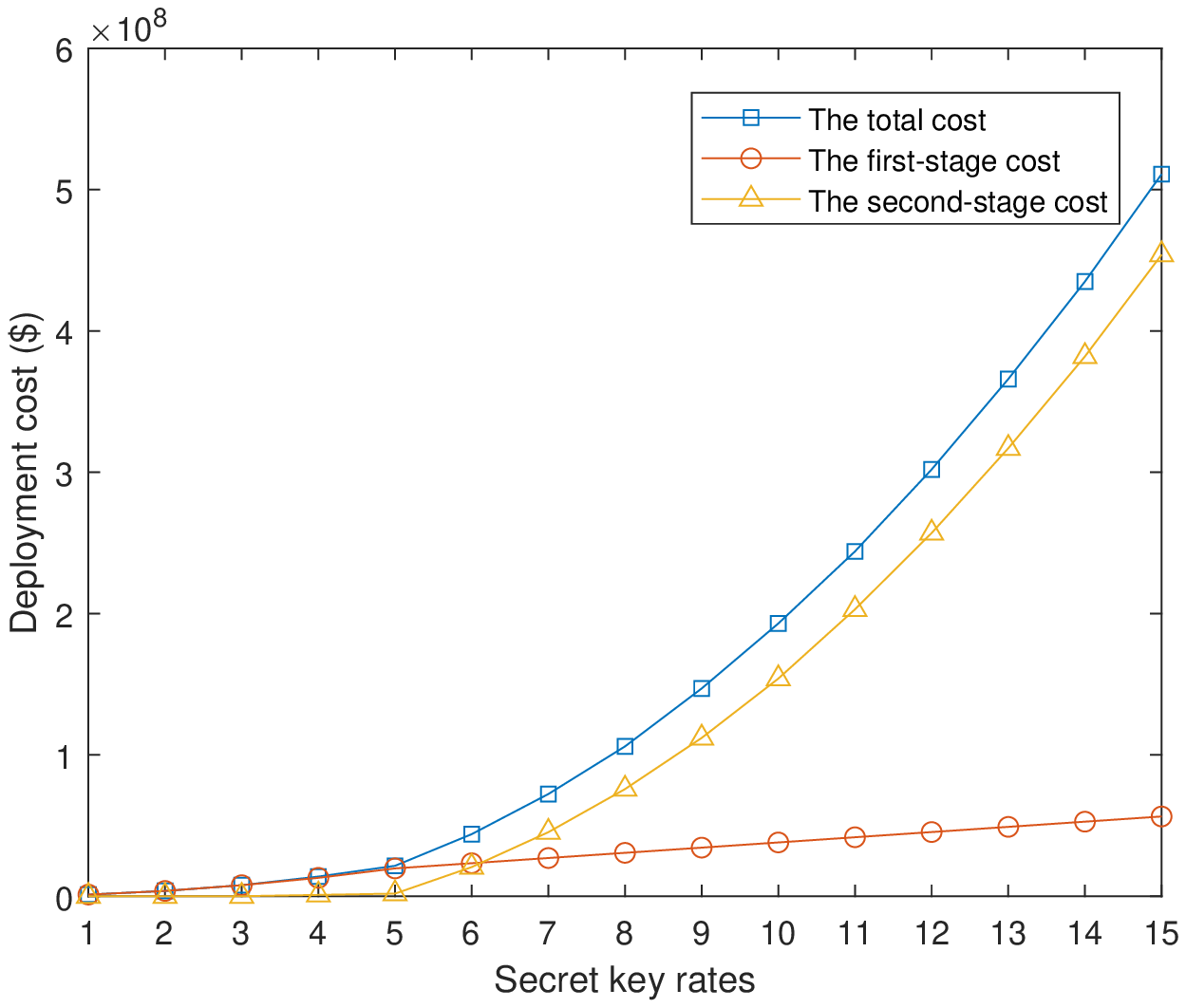}}
 \subfloat[The solution under QKD wavelengths.]{\label{fig:qkd-wave-varied}\includegraphics[width=0.25\textwidth]{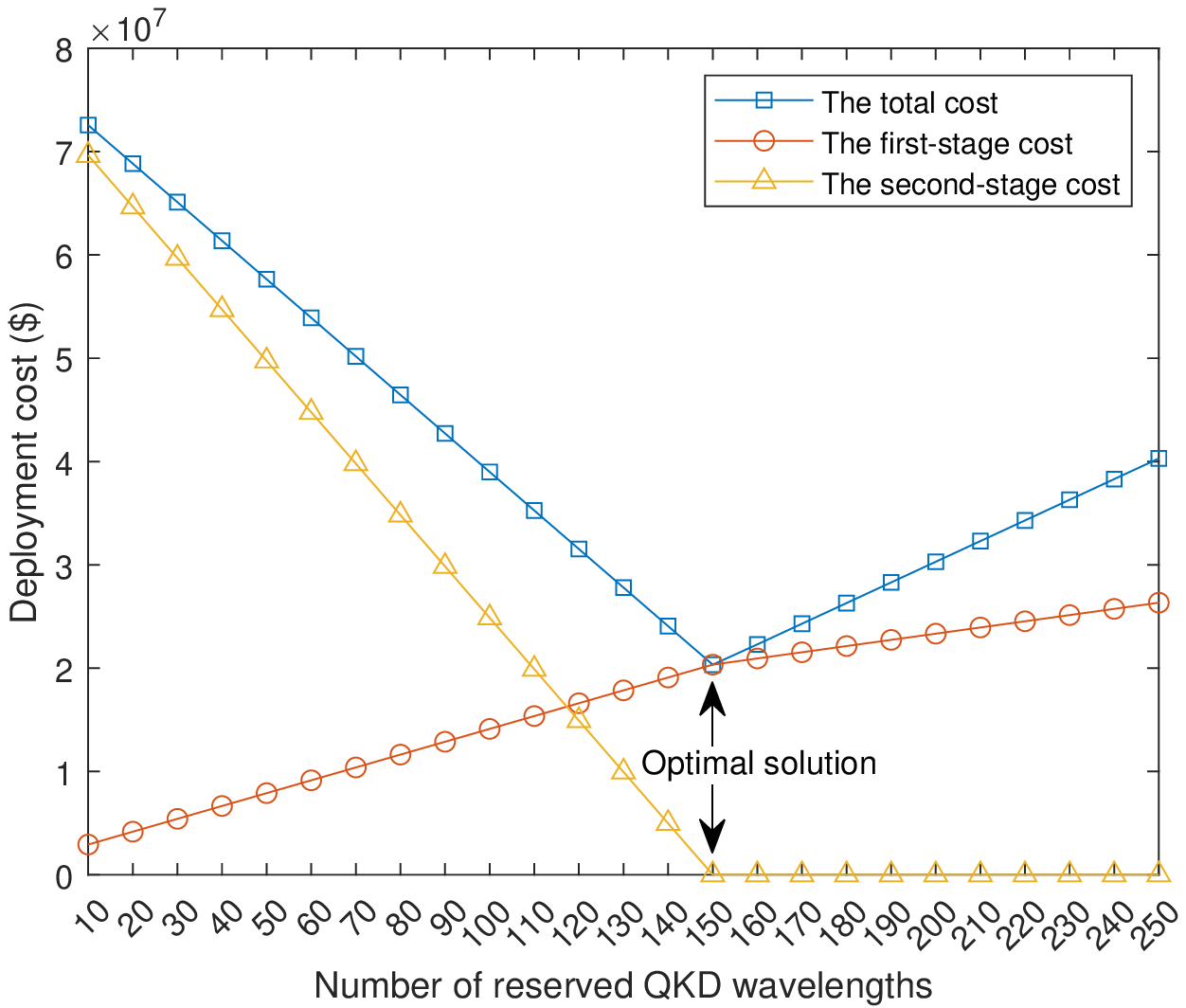}}
 \subfloat[The solution under KM wavelengths.]{\label{fig:km-wave-varied}\includegraphics[width=0.25\textwidth]{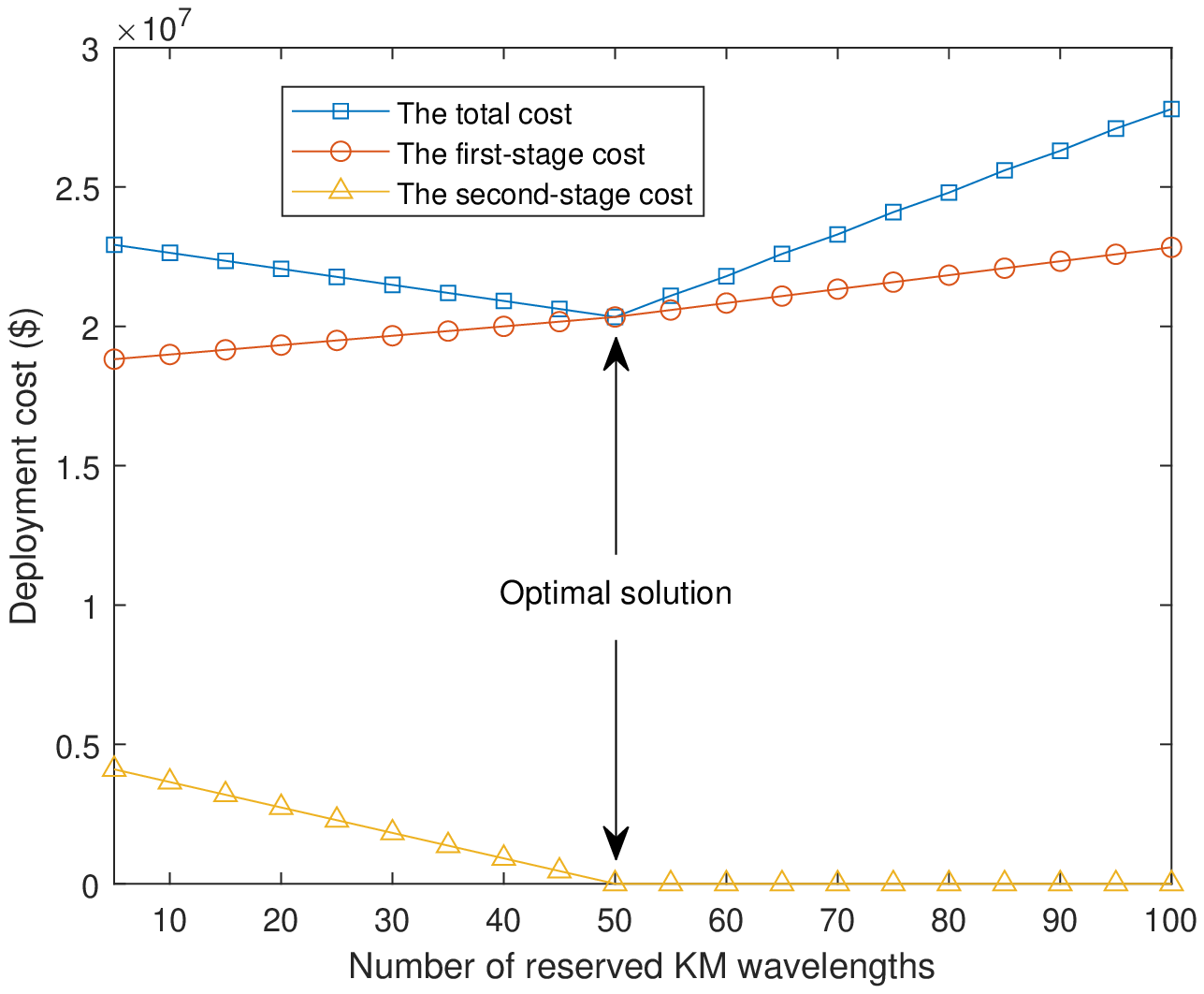}}
 \subfloat[The solution under QKD and KM wavelengths.]{\label{fig:qkd-km-wave-varied}\includegraphics[width=0.25\textwidth]{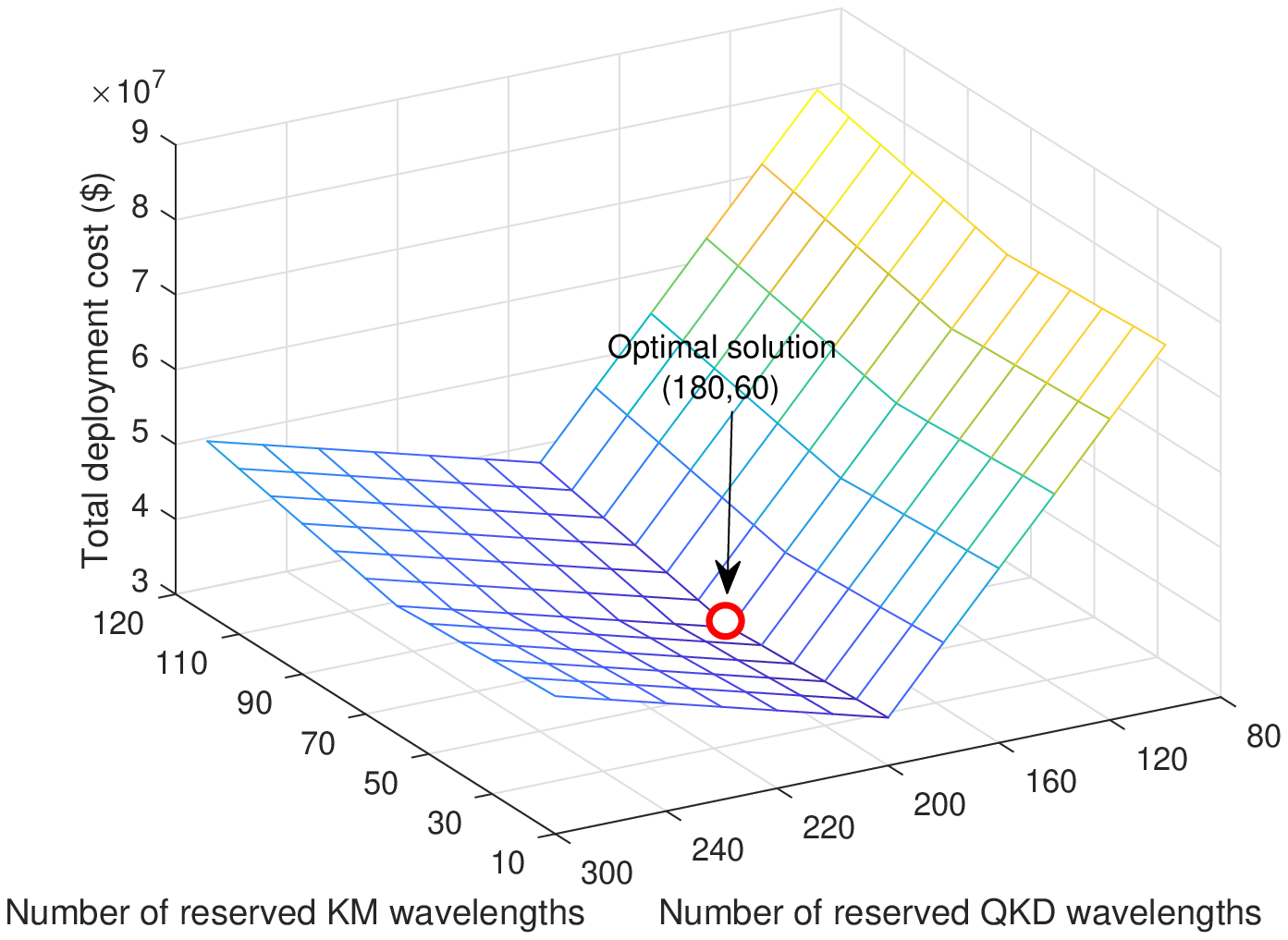}}
 \caption{ (a) The deployment cost in the SP model, (b) The optimal solution under different reserved QKD wavelengths, (c) The optimal solution under different reserved KM wavelengths, and (d) The optimal solution under different reserved QKD and KM wavelengths. }
 \label{fig:deployment-cost}
\end{figure*}
%%%%%%%%%%%%%%%%%%%%%%%%%%%%%%%%%%%%%%%%%%%%%%%%%%%%%%%%%%%%%%%%%%%%%%%%%%%%%%%%%%

In Fig. \ref{fig:deployment-cost}(a), we investigate how the secret-key rate, which is an important demand parameter from users and applications, affects the costs in different stages. Here, the observation is that the second-stage cost increases much faster than the first-stage cost. The reason is that the price of on-demand resources in the second stage is much higher than that of reserved resources in the first stage. In particular, the price of MDI-QTxs and MDI-QRxs in the second stage is high than those in the first stage by four times. Therefore, in the second stage, the price of MDI-QTxs and MDI-QRxs dominates the cost in the second stage since not only the price of MDI-QTxs and MDI-QRxs in the second stage (i.e., the on-demand phase) is much higher than those in the first stage (i.e., the reservation phase) but also the price of MDI-QTxs and MDI-QRxs directly depends on the secret-key rates. 

%In addition, it is straightforward to see that the deployment costs (i.e., the total cost, the first-stage cost, and the second-stage cost) increase when the secret-key rates increase. Moreover, we observe that when the secret-key rates increase, the second-stage cost rises dramatically while the first-stage cost increases more slowly. This is because when the secret-key rate is high, the reserved resources are limited to be utilized. In addition, the cost of on-demand resources dominated by the price of MDI-QTxs and MDI-QRxs in the second stage is higher than that of reserved resources in the first stage and depends on the secret-key rates. From this observation, we can explain that the second-stage cost is a dominant cost when the secret-key rate is high.

In Fig. \ref{fig:deployment-cost}(b), we examine the performance of the SP model in obtaining an optimal solution. In the first stage, we vary the number of reserved QKD wavelengths and fix the number of reserved KM wavelengths. Then, we present the optimal solution obtained by the SP model and the effect of reserved QKD wavelengths on the solution. In Fig. \ref{fig:deployment-cost}(b), when the number of reserved QKD wavelengths increases, the first-stage cost increases significantly. However, the second-stage cost decreases dramatically when the secret-key rates check. The reason is that the QKD wavelength in the on-demand phase (i.e., the second stage) is forced to be minimum by utilizing the cheaper QKD wavelength in the reservation phase (i.e., the first stage). As a result, at 150 QKD wavelengths reserved, the optimal solution can be achieved, and the second-stage cost is 0. This is because the reserved QKD wavelengths meet demands (i.e., secret-key rates), and therefore the on-demand QKD wavelengths in the second stage are not utilized. After this point, the total cost and the first-stage cost increase since it has a penalty cost for an excess of reserved QKD wavelength to be charged. For Fig. \ref{fig:deployment-cost}(b), we can mention that over- and under-provision of QKD wavelengths affect the high total cost significantly. 

In a similar way to the investigation above, in Fig. \ref{fig:deployment-cost}(c), we vary the number of KM wavelengths in the first stage and fix the number of QKD wavelengths to show the optimal solution obtained by the SP model. In Fig. \ref{fig:deployment-cost}(c), clearly, the first-stage cost increases steadily while the second-stage cost decreases. However, the optimal solution is achieved when the number of reserved KM wavelength is 50. The reason is that the reserved KM wavelengths of 50 successfully satisfy the demands. After this point, the total cost and the first-stage cost increase since the penalty cost of an excess of reserved KM wavelengths is charged. Therefore, we can explain that the high cost is affected by not only the reserved QKD wavelengths but also the reserved KM wavelengths. In addition, we vary the number of QKD and KM wavelengths to show the optimal solution obtained by the SP model. In Fig. \ref{fig:deployment-cost}(d), apparently, the optimal deployment cost can be achieved (i.e., the numbers of reserved QKD and KM wavelengths being 180 and 60, respectively) when the number of QKD and KM wavelengths increases.

\begin{comment}
\begin{figure}[htb]
\centering
\captionsetup{justification=centering}
$\begin{array}{c} \epsfxsize=2.5 in \epsffile{./pics/ro-QKD-KM-wavelength.eps} \\
\end{array}$
\caption{QKD and KM wavelength utilization under different secret-key rates.} 
\label{fig:r-o-qkd-km-wavelength}
\end{figure}
\end{comment}

According to the investigation of the secret-key rate, we present the number of QKD and KM wavelengths utilized in reservation and on-demand phases and explain the reason that the second-stage cost increases dramatically. Also, this helps us to understand how the QKD and KM wavelengths affect the second-stage cost. 

\begin{figure*}[htb]
 \centering
 \captionsetup{justification=centering}
 \subfloat[QKD and KM wavelength utilization under different secret-key rates.]{\label{fig:r-o-qkd-km-wavelength}\includegraphics[width=0.25\textwidth]{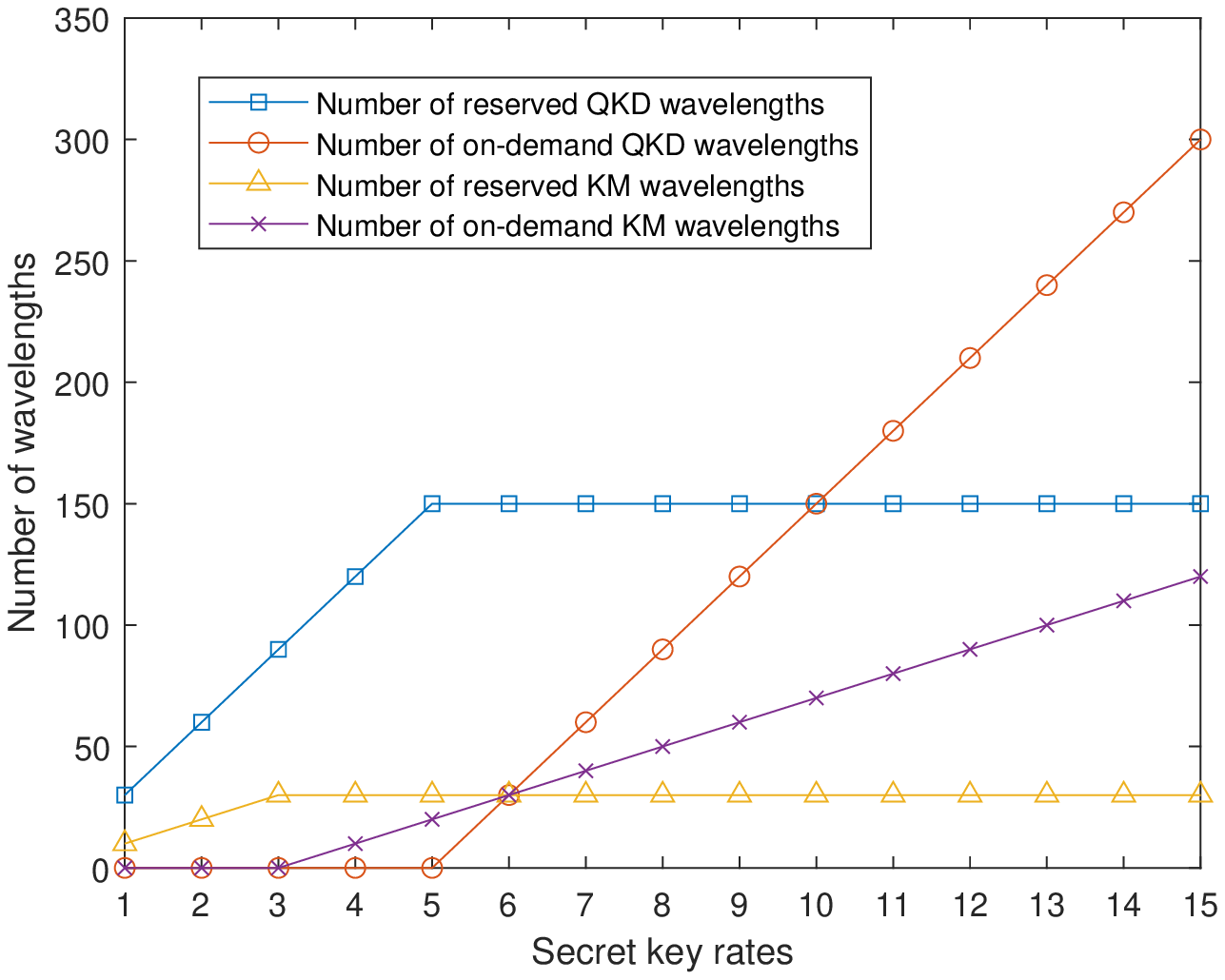}}
 \subfloat[QKD and KM wavelength utilization under different costs.]{\label{fig:qkd-km-link-path}\includegraphics[width=0.25\textwidth]{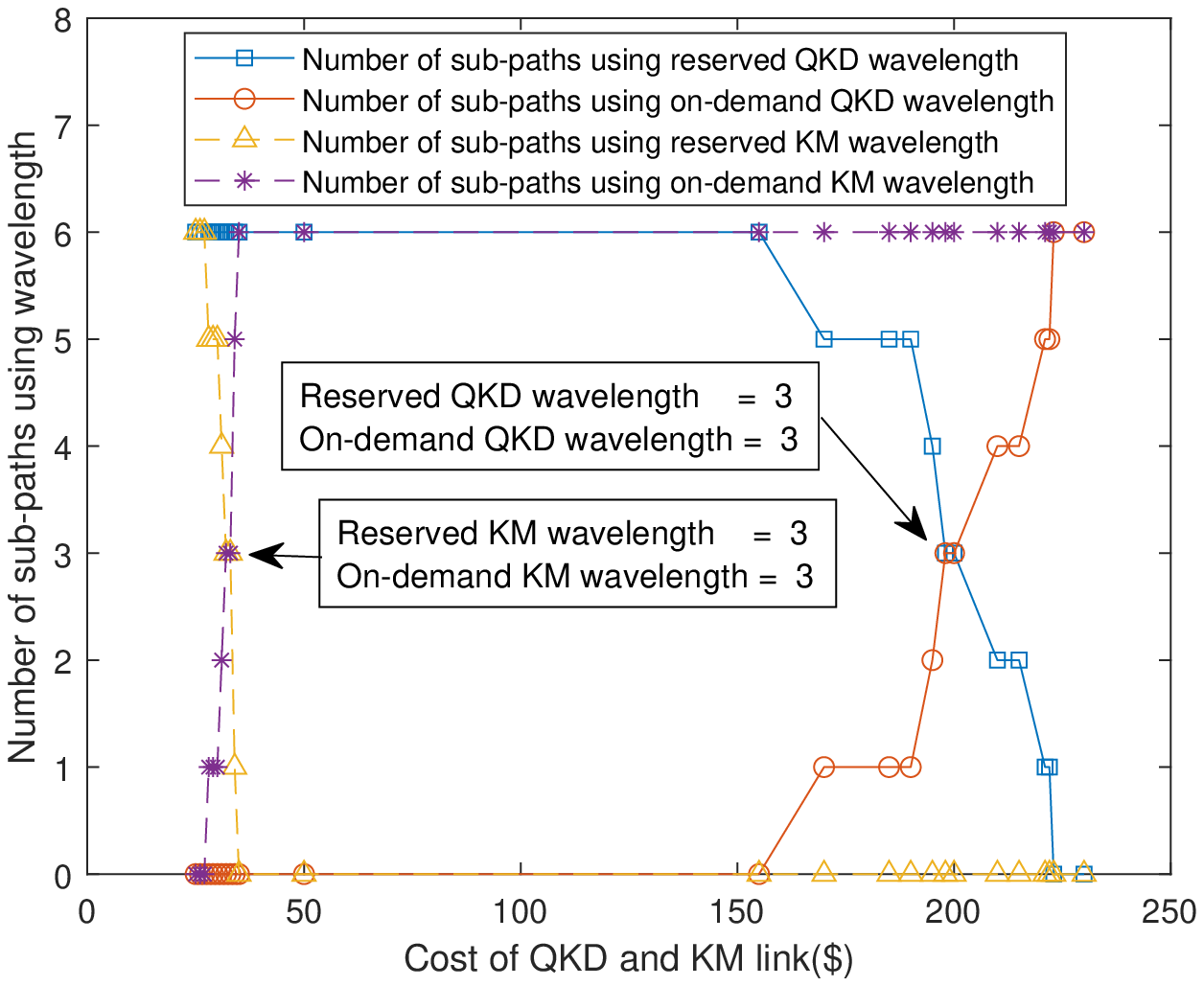}}
 \subfloat[Boundary.]{\label{fig:bounding-sp-model}\includegraphics[width=0.25\textwidth]{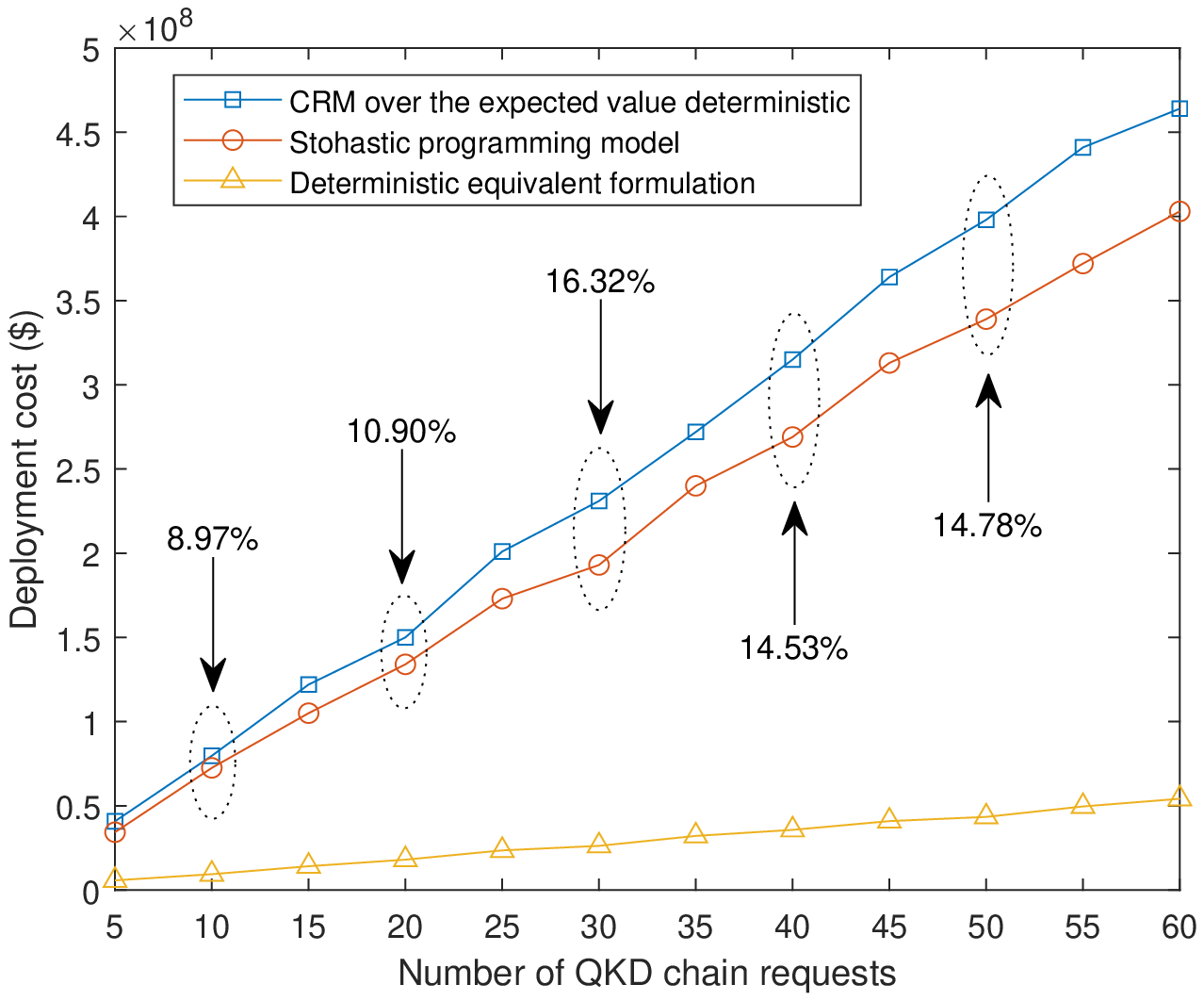}}
 \subfloat[Cost comparison.]{\label{fig:comparison-sp-CO-QBN}\includegraphics[width=0.25\textwidth]{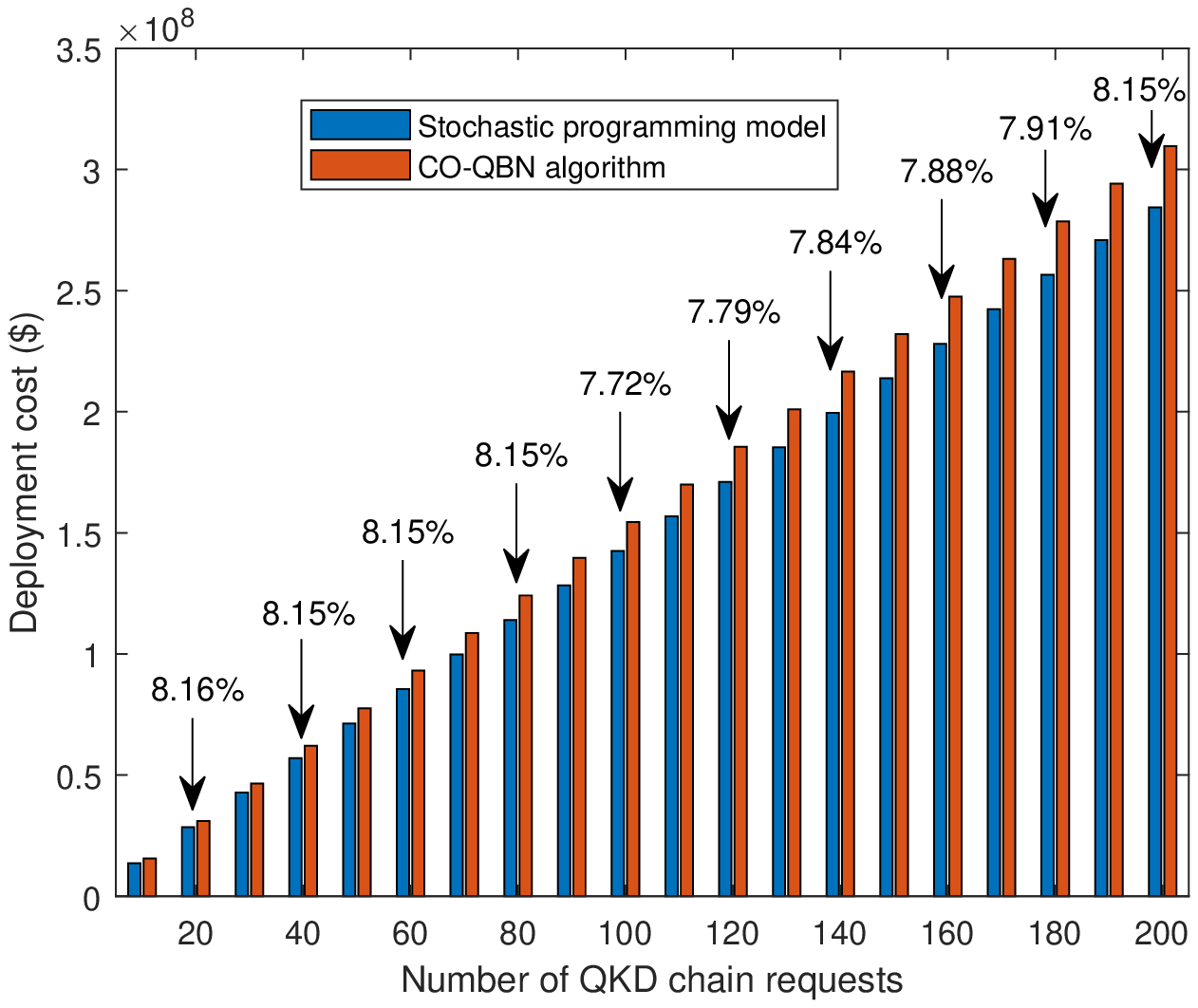}}
 \caption{(a) The numbers of QKD and KM wavelengths in reservation and on-demand phases under different secret-key rates, (b) The numbers of QKD and KM wavelengths in reservation and on-demand phases under the different costs of QKD and KM links, 
 (c) Bounding of the objective value from the SP model, and (d) Cost comparison between the SP model and the CO-QBN algorithm.}
 \label{fig:wavelength-path-bounding-comparison}
\end{figure*}

Figure \ref{fig:wavelength-path-bounding-comparison}(a) illustrates the number of QKD and KM wavelengths (i.e., resources) in reservation and on-demand phases under different secret-key rates. In the reservation phase, the numbers of QKD and KM wavelengths rise steeply until the secret-key rates are 5 and 3 kbps, respectively. This is because the reserved QKD and KM wavelengths are completely utilized at the secret-key rates of 5 and 3 kbps, respectively. After these points, both QKD and KM wavelengths are constant since the capacities of QKD and KM wavelengths are limited. The capacities of QKD and KM wavelengths in the reservation phase are constrained by the QKD service provider. As a result, to satisfy the high secret-key rates, the QKD and KM wavelengths in the on-demand phase are more utilized. In the on-demand phase, the numbers of QKD and KM wavelengths start rising dramatically at 5 and 3 kbps, respectively, due to the limited capacities of QKD and KM wavelengths in the reservation phase.

\begin{comment}
\begin{figure}[htb]
\centering
\captionsetup{justification=centering}
$\begin{array}{c} \epsfxsize=2.5 in \epsffile{./pics/qkd-km-link4.eps} \\
\end{array}$
\caption{QKD and KM wavelength utilization under different costs.} 
\label{fig:qkd-km-link-path}
\end{figure}
\end{comment}

Figure \ref{fig:wavelength-path-bounding-comparison}(b) illustrates sub-path utilization according to the QKD-SIC chain request (i.e., from a source to a destination) of QKD and KM wavelength in reservation and on-demand phases under the different costs of QKD and KM links. The full path in Fig. \ref{fig:wavelength-path-bounding-comparison}(b) is $1 \rightarrow  6 \rightarrow 9 \rightarrow 12 \rightarrow 16 \rightarrow 22 \rightarrow 23$, which is consisted of 6 sub-paths. Source and destination nodes are 1 and 23, respectively. In Fig. \ref{fig:wavelength-path-bounding-comparison}(b), it is straightforward to mention that when the cost of QKD and KM links increases, the number of utilized sub-paths of both QKD and KM wavelengths in the reservation phase changes to that of both QKD and KM wavelengths in the on-demand phase. For example, with QKD and KM link cost $< 50\$$, the sub-paths of KM wavelengths in reservation and on-demand phases are utilized. However, with QKD and KM link cost = 50\$, the sub-paths of KM wavelength in the on-demand phase are completely utilized (i.e., six sub-paths). To obtain the optimal deployment cost, the sub-path utilization in the reservation phase is completely changed to the sub-path utilization in the on-demand phase by the SP model. 

%%%%%%%%%%%%%%%%%%%%%%%%%%%%%%%%%%%%%%%%%%%%%%%%%%%%%%%%%%%%%%%%%%%%%%%%%
%%%%%%%%%%%%%%%%%%%%%%%%%%%%%%%%%%%%%%%%%%%%%%%%%%%%%%%%%%%%%%%%%%%%%%%%%

\subsubsection{Performance Evaluation Under Various Parameters}
\label{subsubsec:performance-ev-various-parameters}

We first evaluate the performance of the SP model by comparing it with the \emph{cooperative resource management (CRM) over the expected value deterministic} and the \emph{deterministic equivalent formulation}. In the case of CRM over the expected value deterministic, the secret-key rates (i.e., demands) in the first stage are considered to be the expected demands. On the other hand, the secret-key rates are considered as the exact demands in the case of the deterministic equivalent formulation. The former formulation is used to find the values of the upper bound, while the latter formulation is used to find the values of the lower bound. 
%These formulations are used to find the values of upper and lower bounds.

\begin{comment}
\begin{figure}[htb]
\centering
\captionsetup{justification=centering}
$\begin{array}{c} \epsfxsize=2.5 in \epsffile{./pics/boundary-comparison1.eps} \\
\end{array}$
\caption{Bounding of the objective values from the SP model.} 
\label{fig:bounding-sp-model}
\end{figure}
\end{comment}

In Fig. \ref{fig:wavelength-path-bounding-comparison}(c), the CRM over the expected value deterministic and deterministic equivalent formulation yield the solutions that are the upper bound and lower bound of the SP model, respectively. We observe that the gap to the upper bound is sufficiently small. This result implies that using the expected value of secret-key rates can achieve a satisfactory solution (i.e., close to the optimal solution of the SP model). For example, when the QKD-SIC chain requests are 10, which has the slight upper bound (i.e., 8.97\%) shown in Fig. \ref{fig:wavelength-path-bounding-comparison}(c), we can obtain a good solution by using CRM over the expected value deterministic instead of the SP model due to low computational complexity. 

\begin{comment}
\begin{figure}[htb]
\centering
\captionsetup{justification=centering}
$\begin{array}{c} \epsfxsize=2.5 in \epsffile{./pics/comparison-bar-graph1.eps} \\
\end{array}$
\caption{Cost comparison between the SP model and CO-QBN algorithm.} 
\label{fig:comparison-sp-CO-QBN}
\end{figure}
\end{comment}

Then, we compare the SP model with the CO-QBN algorithm \cite{y-cao-hybrid-trusted2021}. Figure \ref{fig:wavelength-path-bounding-comparison}(d) illustrates a performance comparison of the SP model and CO-QBN algorithm by varying the number of QKD-SIC chain requests. In Fig. \ref{fig:wavelength-path-bounding-comparison}(d), both the SP model and CO-QBN algorithm can decrease the deployment cost when the QKD-SIC chain requests increase. However, in comparison with the CO-QBN algorithm, the SP model can significantly achieve a lower deployment cost under different QKD-SIC chain requests. For example, with the QKD-SIC chain requests of 60, the SP model can reduce the deployment cost compared with the CO-QBN algorithm by 8.15\%. In addition, we observe that the performance of the SP model compared with the CO-QBN algorithm under different QKD-SIC chain requests is quite stable (i.e., 8.21\%). Therefore, we can claim that the SP model significantly outperforms the CO-QBN algorithm.   

\subsubsection{Impact of Available QKD and KM wavelengths}
\label{subsubsec:impact-of-available-qkd-km-wave}

\begin{figure*}[htb]
 \centering
 \captionsetup{justification=centering}
 \subfloat[Total deployment cost under different QKD and KM wavelengths shared by provider 2.]{\label{fig:qkm-km-shared-provider2}\includegraphics[width=0.33\textwidth]{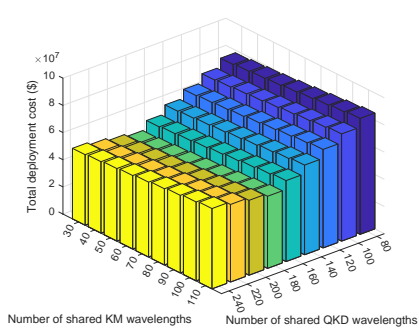}}
 \subfloat[Cost of providers 1 and 2 under different QKD wavelengths.]{\label{fig:co-no-qkd-cooperation}\includegraphics[width=0.33\textwidth]{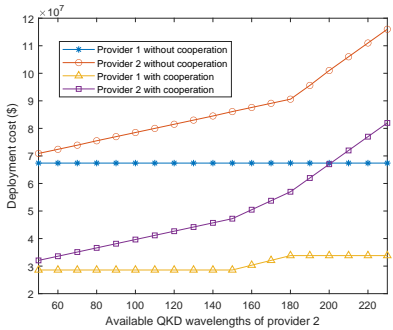}}
 \subfloat[Cost of providers 1 and 2 under different KM wavelengths.]{\label{fig:co-no-km-cooperation}\includegraphics[width=0.33\textwidth]{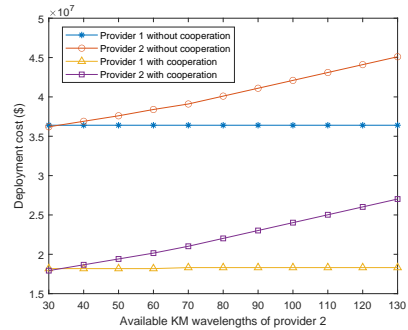}}
 \caption{(a) The deployment cost under different number of shared QKD and KM wavelengths of provider 2, (b) Costs of providers 1 and 2 with and without cooperation under different available QKD wavelengths of provider 2, and (c) Costs of providers 1 and 2 with and without cooperation under different available KM wavelengths of provider 2.}
 \label{fig:cost-coop1-2}
\end{figure*}

We consider a situation that the QKD service providers 1 and 2 cooperate to share the QKD and KM wavelengths. Figure \ref{fig:cost-coop1-2}(a) illustrates the total deployment cost when provider 2 shares the number of QKD and KM wavelengths. It is clear to explain that, with the high demands, the total deployment cost increases if provider 2 performs over-sharing wavelengths and under-sharing wavelengths. Figures \ref{fig:cost-coop1-2}(b) and (c) illustrate the optimal deployment cost obtained from the SP model when provider 2 increases the number of shared QKD and KM wavelengths. The main observation is that the providers experience lower costs when they are in cooperation. This is due to the fact that the providers can utilize the available resources of each other. As illustrated in Figs. \ref{fig:cost-coop1-2}(b) and \ref{fig:cost-coop1-2}(c), when the QKD and KM wavelengths of provider 2 increase, provider 1 receives a benefit from using the extra QKD and KM wavelengths of provider 2 and provider 2 can utilize QKD and KM wavelengths shared in the resource pool by provider 1. On the other hand, without cooperation, the QKD and KM wavelength resources cannot be shared. Hence, increasing the available QKD and KM wavelengths of provider 2 does not have an effect on the deployment cost of provider 1. Another observation is that, with cooperation, the cost of provider 1 can slightly increase when the QKD and KM wavelengths of provider 2 increase. This is because provider 1 does not contribute more to the QKD and KM wavelength resource pool. Therefore, the cost-sharing of provider 1 increases when provider 2 contributes more resources.  

\subsubsection{Cooperation Formulation and Impact of Cooperation Cost}
\label{subsubsec:cooperation-formulation-cost}

%%%%%%%%%%%%%%%%%%%%%%%% In case of 1N, 2N, 3N %%%%%%%%%%%%%%%%%%%%%%%%%%%%%%%%%%%%%%%%%%%%%%%
\begin{table*}[htb] \footnotesize \caption{Payoff matrix for three coalition game}
\label{table:shapley-value-coalition}
\centering
\scalebox{0.9}{\begin{tabular}{|c|c|c|c|c|c|c|}\hline
\multicolumn{7}{|c|}{\bf{The Shapley value}} \\\hline
&		\multicolumn{3}{|c|}{QKD wavelength pool} & \multicolumn{3}{|c|}{ KM wavelength pool}   \\\hline
Cooperation & Provider     & Provider	  &	Provider	  &  Provider	   & Provider	& Provider	  \\
structure	& 1  	       & 2	          &	3	          &  1 	           & 2          &  3   \\\hline
$\mathbb{C}_{1} = \{\{1\}, \{2\},\{3\}\}$ & 3,271,643.12 &2,998,812.40	&2,725,981.68   & 35,647,210.00	 &35,260,720.00	&35,131,890.00   \\\hline
$\mathbb{C}_{2} = \{\{1,2\}, \{3\}\} $	  & {\bf 2,562,990.84}* &{\bf 2,890,160.12}*	&{\bf 2,725,981.68}*   & 32,199,190.00	 &37,227,700.00	&35,131,890.00   \\\hline
$\mathbb{C}_{3} = \{\{1,3\}, \{2\}\} $ 	  & 2,562,990.84 &2,998,812.40	&3,217,329.40   & 32,263,605.00	 &35,260,720.00	&40,773,285.00   \\\hline
$\mathbb{C}_{4} = \{\{2,3\}, \{1\}\}$	  & 3,271,643.12 &3,026,575.48  &3,353,744.76   & 35,647,210.00	 &37,485,360.00	&40,966,530.00   \\\hline
$\mathbb{C}_{5}=  \{\{1,2,3\}\} $	      & 2,108,660.56 &2,572,245.20  &2,899,414.48   & {\bf 26,300,931.67}*	 &{\bf 31,522,686.67}*	&{\bf 35,068,271.67}*   \\\hline 
\end{tabular}}
\end{table*}	
%%%%%%%%%%%%%%%%%%%%%%%%%%%%%%%%%%%%%%%%%%%%%%%%%%%%%%%%%%%%%%%%%%%%%%%%%%%%%%%%%%%%%%%%%%%%%

We evaluate the cooperation formation behavior of the three providers and then scrutinize the cooperation costs which comprise shared QKD wavelength (i.e., $C^{\mathrm{qkd}}(s)$), shared KM wavelength (i.e., $C^{\mathrm{kmw}}(s)$), and cooperation (i.e., $C^{\mathrm{coc}}_s$). Table \ref{table:shapley-value-coalition} illustrates the Shaplye values (i.e., the cost of each provider), which is $\vartheta_{s}(v)$ = $\vartheta_{s}(v)$ + $C^{\mathrm{qkd}}(s)$ +  $C^{\mathrm{kmw}}(s)$ + $C^{\mathrm{coc}}_s$, for the QKD and KM wavelength resource allocation. With cost management and QKD and KM wavelength allocation based on the SP model, the stable coalition structure of the QKD wavelength pool is obtained to be $\mathbb{C}_{2}^{*}$ (i.e., providers 1 and 2 cooperate). $\mathbb{C}_{2}^{*}$ is a stable coalition structure since providers 1 and 2 achieve the lowest cost, while provider 3 does not have a better choice to decrease its cost. In addition, the stable coalition structure of KM wavelength pool is obtained to be $\mathbb{C}_{5}^{*}$ (i.e., providers 1, 2, and 3 cooperate). $\mathbb{C}_{5}^{*}$ is a stable coalition structure since providers 1, 2, and 3 achieve a lower cost compared with other coalition structures.  

\begin{figure*}[htb]
 \centering
 \captionsetup{justification=centering}
 \subfloat[Transition cooperation structure under different cost of QKD wavelength.]{\label{fig:transition-qkd-wavelength}\includegraphics[width=0.34\textwidth]{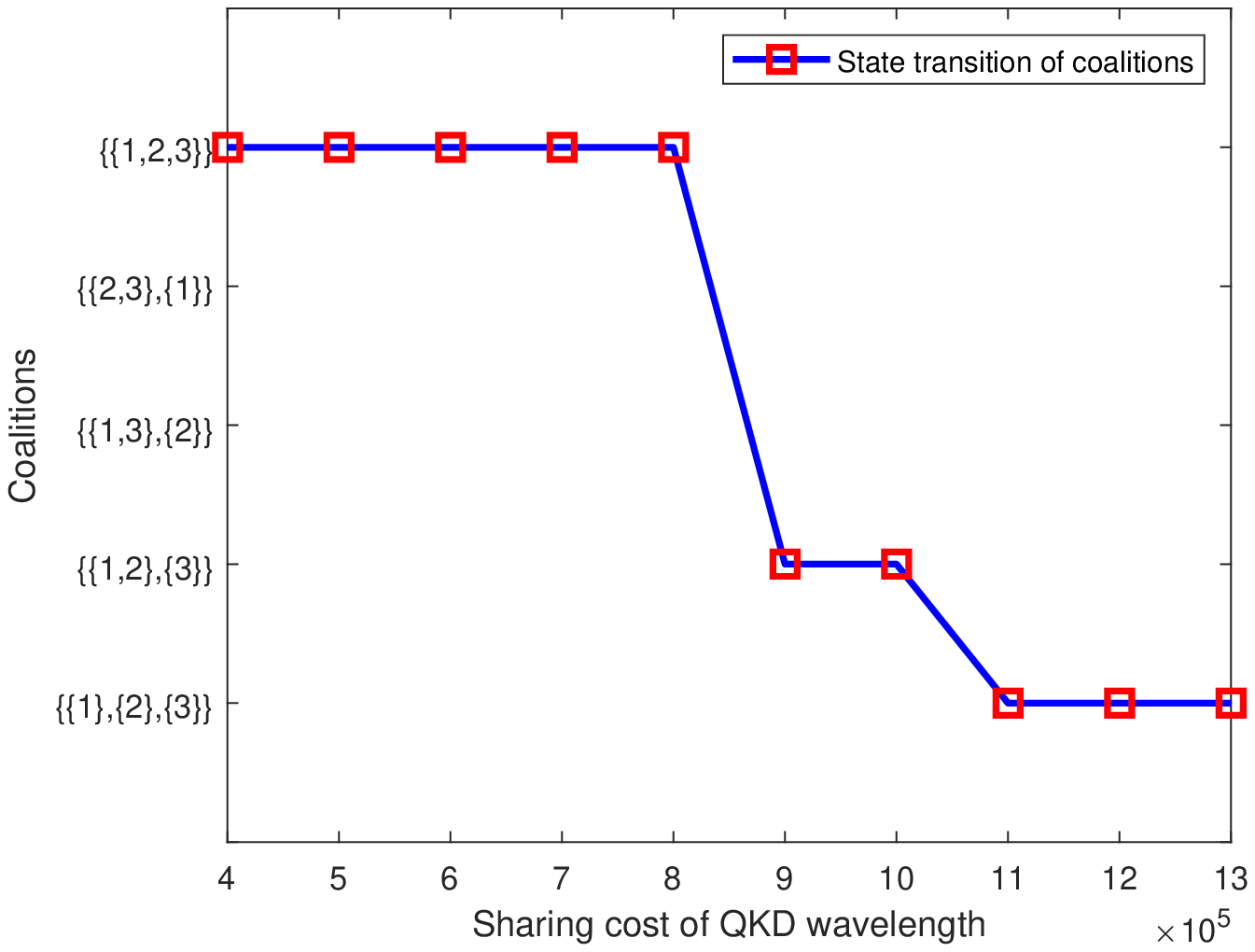}}
 \subfloat[Transition cooperation structure under different cost of KM wavelength.]{\label{fig:transition-km-wavelength}\includegraphics[width=0.34\textwidth]{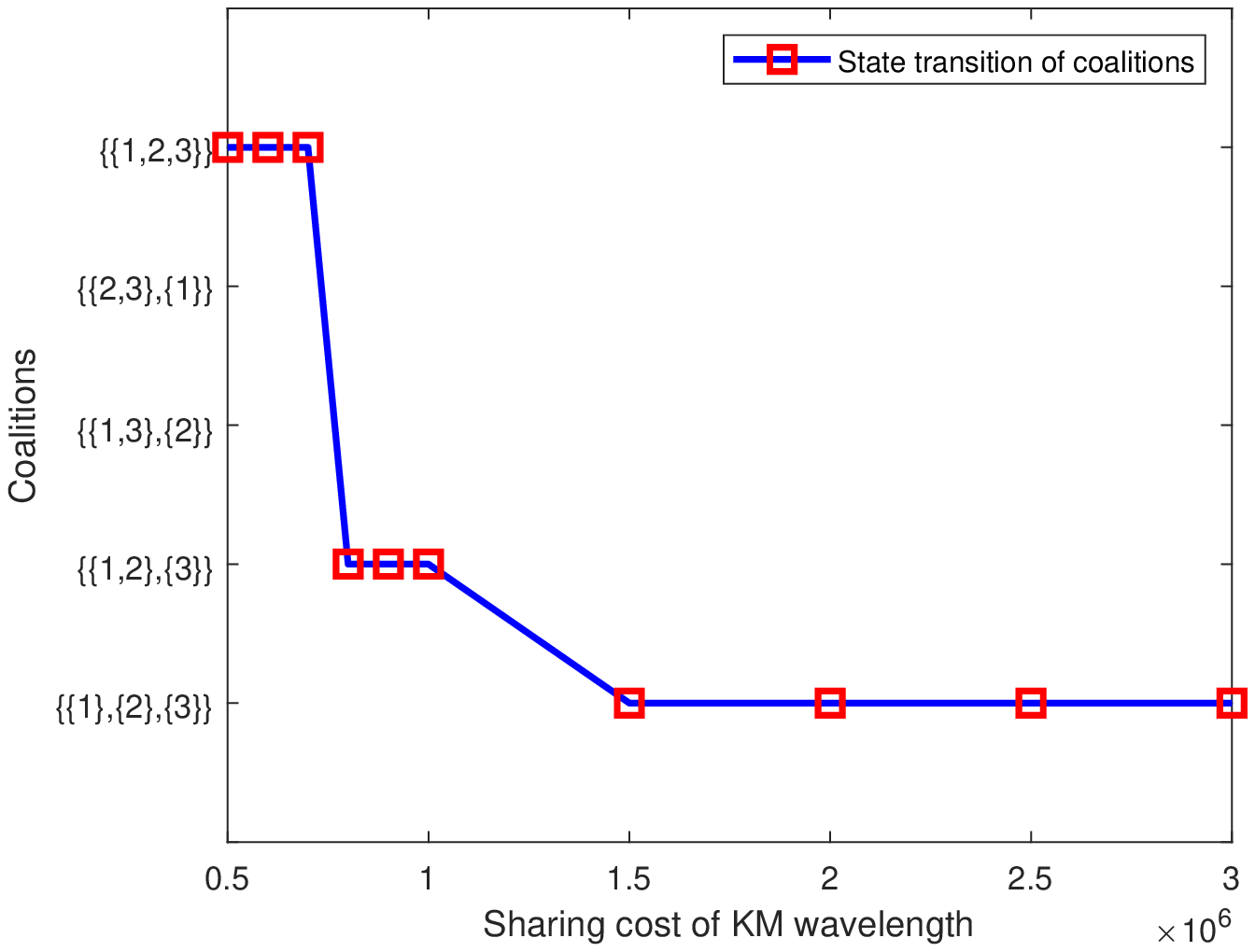}}
 \subfloat[Transition cooperation structure under different cooperation cost of each provider.]{\label{fig:transition-coopration-cost}\includegraphics[width=0.34\textwidth]{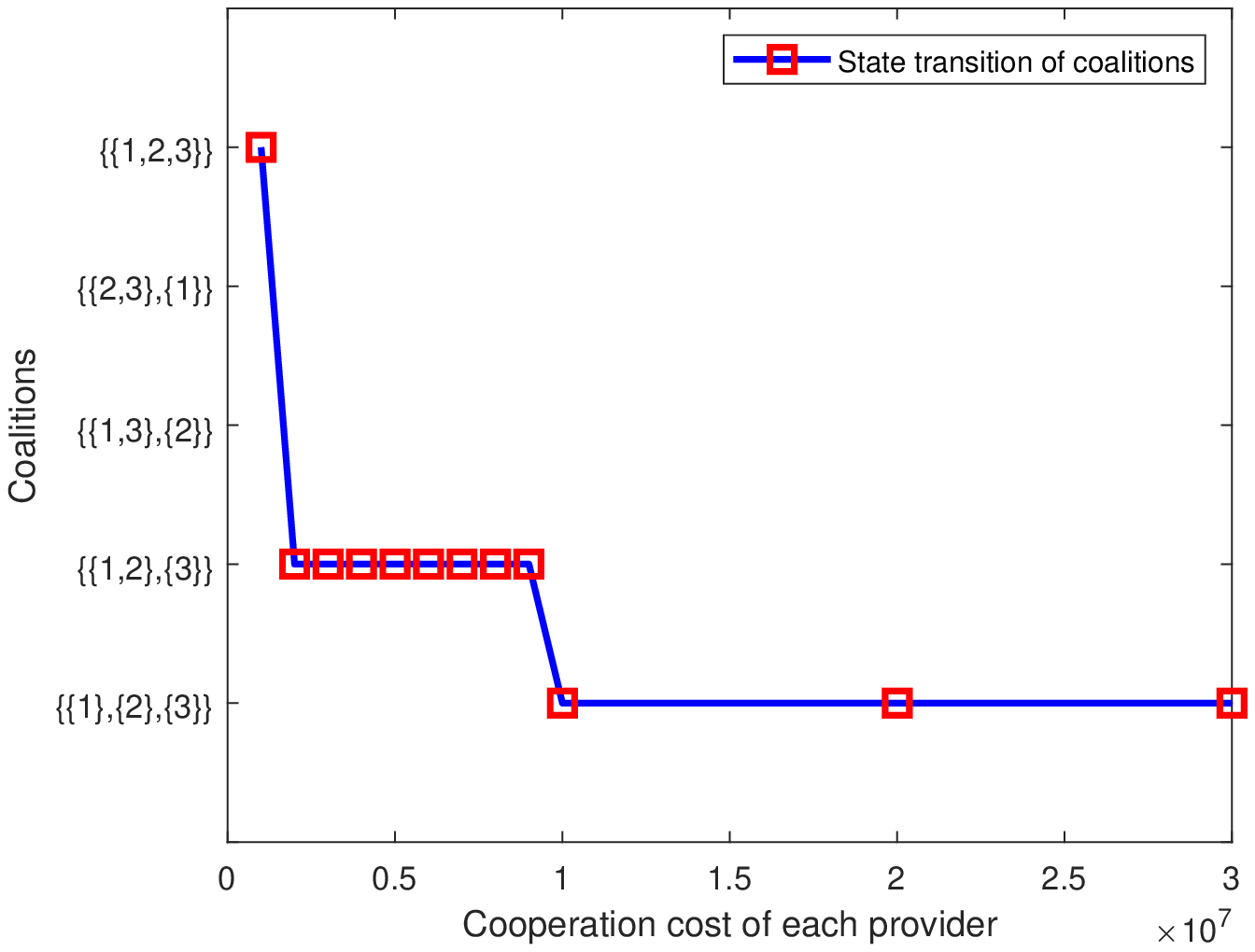}}
 \caption{(a) The transition cooperation structure under different costs of QKD wavelength, (b) KM wavelength, and (c) cooperation of each provider.}
 \label{fig:state-transition}
\end{figure*}

We investigate how the costs of parameters (i.e., shared QKD and KM wavelengths, and cooperation) affect the stable cooperation structure. In Figs. \ref{fig:state-transition}(a), \ref{fig:state-transition}(b), and \ref{fig:state-transition}(c), the major observation is that the cooperation structure will move towards $\{\{1\}, \{2\}, \{3\}\}$ which means that providers work separately when the costs of shared QKD wavelength (i.e., $C^{\mathrm{qkd}}(s)$), shared KM wavelength (i.e., $C^{\mathrm{kmw}}(s)$), and cooperation (i.e., $C^{\mathrm{coc}}_s$) increase. The reason is that, without cooperation, the providers mainly utilize the resources (i.e., QKD and KM wavelengths) in reservation and on-demand phases to obtain the minimum cost. When the costs increase, it is not worth for the provider to cooperate and increase their resources to accommodate demands. Alternatively, it is more beneficial for the provider to work separately.

\begin{figure}[htb]
\centering
\captionsetup{justification=centering}
$\begin{array}{c} \epsfxsize=3.3 in \epsffile{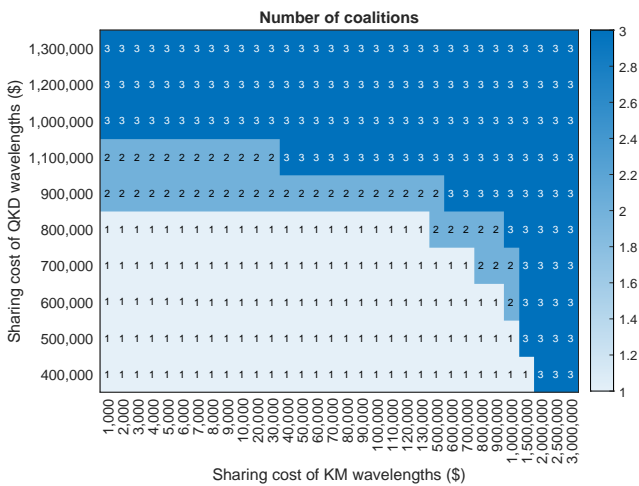} \\
\end{array}$
\caption{Stable coalitions under different sharing costs of QKD and KM wavelengths.} 
\label{fig:state-transition-coalitions}
\end{figure}

Figure \ref{fig:state-transition-coalitions} illustrates the adaptation of the stable coalitions under different sharing costs of QKD and KM wavelengths. With a large number of secret-key rates (i.e., demands), each provider tends to move to the coalition structure $\mathbb{C}_{5}=  \{\{1,2,3\}\} $ when the sharing cost of QKD and KM wavelengths is low. For example, with the sharing cost of QKD and KM wavelengths being 600,000\$ and 10,000\$, respectively, the stable coalition is $\mathbb{C}_{5}=  \{\{1,2,3\}\} $ which is indicated by ``1'' in Fig. \ref{fig:state-transition-coalitions}. This is because each provider obtains the lowest cost, and the resource can be better utilized with more providers in the coalition. However, when the sharing cost of QKD and/or KM wavelengths increases, the stable coalition tends to move to the coalition $\mathbb{C}_{1} = \{\{1\}, \{2\},\{3\} \}$ which is indicated by ``3'' in Fig. \ref{fig:state-transition-coalitions}. This is due to the fact that each provider obtains a high cost when they cooperate to share their QKD and/or KM wavelengths. As a result, there are no QKD and KM wavelength pools, and each provider separately supports the demands.  

\section{Conclusions}
\label{sec:conclusion}
%% QKD-SIC 
In the proposed QKD-SIC system, edge devices will need security requirements in QKD nodes. Therefore, the QKD service providers must provide QKD resources to achieve the minimum deployment cost. We have proposed a decision-making scheme for QKD service providers in QKD-based SIC. The proposed scheme comprises methods for resource allocation to secure semantic information transmission of edge devices, cost management, and cooperation among QKD service providers. In particular, the QKD service providers can cooperate and establish the QKD resource pool. For the resource allocation, we have applied the two-stage stochastic programming to resolve the solutions of the QKD service providers to support edge devices to transmit the semantic information that utilizes resources in the resource pool. For cost management, we have incorporated Shapley values to share the deployment cost that is received from the resource pool with cooperative QKD service providers in a fair and interpretable manner. For the coalition formation among QKD service providers, we have achieved the cooperation equilibrium solutions for the cooperative QKD service providers who contribute to the resource pool.

%\bibliographystyle{ieeetr}
%\bibliography{ref}

\begin{thebibliography}{10}


\bibitem{cao2022evolution}
Y. Cao, Y. Zhao, Q. Wang, J. Zhang, S. X. Ng and L. Hanzo, ``The Evolution of Quantum Key Distribution Networks: On the Road to the Qinternet,'' in {\em IEEE Communications Surveys \& Tutorials}, vol. 24, no. 2, pp. 839-894, 2022.

\bibitem{a-s-cacciapuoti-quantum-2020}
A. S. Cacciapuoti, M. Caleffi, F. Tafuri, F. S. Cataliotti, S. Gherardini, and G. Bianchi, ``Quantum internet: Networking challenges in distributed quantum computing,'' {\em IEEE Network}, vol. 34, no. 1, pp. 137-143, 2020.

\bibitem{a-s-cacciapuoti-entanglement-2020}
A. S. Cacciapuoti, M. Caleffi, R. Van Meter, and L. Hanzo, ``When entanglement meets classical communications: Quantum teleportation for the quantum internet,'' {\em IEEE Transactions on Communications}, vol. 68, no. 6, pp. 3808-3833, 2020.

\bibitem{j-illiano-quantum-internet-2022}
J. Illiano, M. Caleffi, A. Manzalini, and A. S. Cacciapuoti, ``Quantum internet protocol stack: A comprehensive survey,'' {\em Computer Networks}, vol. 213, p. 109092, 2022.

\bibitem{ladd2010quantum}
T. D. Ladd, F. Jelezko, R. Laflamme, Y. Nakamura, C. Monroe, and J. L. O'Brien, ``Quantum computers,'' {\em nature}, vol. 464, no. 7285, pp. 45-53, 2010.

\bibitem{wootters1982single}
W. K. Wootters and W. H. Zurek, ``A single quantum cannot be cloned,'' {\em Nature}, vol. 299, no. 5886, pp. 802-803, 1982.

\bibitem{shannon1949communication}
C. E. Shannon, ``Communication theory of secrecy systems,'' {\em The Bell system technical journal}, vol. 28, no. 4, pp. 656-715, 1949.

\bibitem{you2021towards}
X. You, C.-X. Wang, J. Huang, X. Gao, Z. Zhang, M. Wang, Y. Huang, C. Zhang, Y. Jiang, J. Wang, {\em et al.}, ``Towards 6G wireless communication networks: vision, enabling technologies, and new paradigm shifts''. {\em Sci. China Inf. Sci.}, vol. 64, no. 1, pp. 1-74, 2021.

\bibitem{x-luo-semantic-communications2022}
X. Luo, H.-H. Chen, and Q. Guo, ``Semantic communications: Overview, open issues, and future research directions,'' {\em IEEE Wireless Communications}, vol. 29, no. 1, pp. 210-219, 2022.

\bibitem{y-wanting-semantic-communi2022}
W. Yang, H. Du, Z. Liew, W. Y. B. Lim, Z. Xiong, D. Niyato, X. Chi, X. S. Shen, and C. Miao, ``Semantic communications for 6G future internet: Fundamentals, applications, and challenges,'' {\em arXiv:2207.00427}, 2022.

\bibitem{q-zhijin-semantic-comm2022}
Z. Qin, X. Tao, J. Lu, W. Tong, and G. Y. Li, ``Semantic communications: Principles and challenges,'' {\em arXiv:2201.01389}, 2022.

\bibitem{j-blesswin-enhanced-semantic2019}
A. J. Blesswin, C. Raj, R. Sukumaran, and M. G. Selva, ``Enhanced semantic visual secret sharing scheme for the secure image communication,'' {\em Multimedia Tools and Application}, vol. 79, pp. 17057-17079, 2019.

\bibitem{m-xu-stochastic-resource2022}
M. Xu, W. C. Ng, D. Niyato, H. Yu, C. Miao, D. I. Kim, and S. Shen, ``Stochastic resource allocation in quantum key distribution for secure federated learning,'' in {\em Proceedings of IEEE GLOBECOM}, 2022.

\bibitem{kaewpuang2022-adaptive}
R. Kaewpuang, M. Xu, D. Niyato, H. Yu, and Z. Xiong, ``Adaptive Resource Allocation in Quantum Key Distribution (QKD) for Federated Learning'', {arXiv:2208.11270}, 2022.

\bibitem{zaarour2022openpubsub}
T. Zaarour, A. Bhattacharya, and E. Curry, ``OpenPubSub: Supporting large semantic content spaces in peer-to-peer publish/subscribe systems for the internet of multimedia things,'' {\em IEEE Internet of Things Journal}, 2022.


\bibitem{kim2020compiler}
B. Kim, S. Heo, J. Lee, S. Jeong, Y. Lee, and H. Kim, ``Compiler assisted semantic-aware encryption for efficient and secure serverless computing,'' {\em IEEE Internet of Things Journal}, vol. 8, no. 7, pp. 5645-5656, 2020.

\bibitem{hu2019things2vec}
L. Hu, G. Wu, Y. Xing, and F. Wang, ``Things2vec: Semantic modeling in the internet of things with graph representation learning,'' {\em IEEE Internet of Things Journal}, vol. 7, no. 3, pp. 1939-1948, 2019.

\bibitem{qiu2020mobile}
G. Qiu, D. Guo, Y. Shen, G. Tang, and S. Chen, ``Mobile semantic-aware trajectory for personalized location privacy preservation,'' {\em IEEE Internet of Things Journal}, vol. 8, no. 21, pp. 16165-16180, 2020.

\bibitem{w-c-ng-stochastic-approach2022}
W. C. Ng, H. Du, W. Y. B. Lim, Z. Xiong, D. Niyato, and C. Miao, ``A Stochastic Approach for Semantic-enabled Transmission in the Metaverse,'' To be submitted.

\bibitem{n-h-chu-metaslicing2022}
N. H. Chu, D. T. Hoang, D. N. Nguyen, K. T. Phan, and E. Dutkiewicz, ``Metaslicing: A novel resource allocation framework for metaverse,'' {\em arXiv:2205.11087}, 2022.

\bibitem{yan-lei-resource-allocation2022}
L. Yan, Z. Qin, R. Zhang, Y. Li, and G. Y. Li, ``Resource allocation for text semantic communications,'' {\em IEEE Wireless Communications Letters}, vol. 11, no. 7, pp. 1394-1398, 2022.

\bibitem{w-zhenzi-semantic-comm2021}
Z. Weng and Z. Qin, ``Semantic communication systems for speech transmission,'' {\em IEEE Journal on Selected Areas in Communications}, vol. 39, no. 8, pp. 2434-2444, 2021.

\bibitem{ma2022equilibrium}
W. Ma, B. Chen, L. Liu, H. Chen, W. Shao, M. Gao, J. Wu, and P.-H. Ho, ``Equilibrium allocation approaches of quantum key resources with security levels in qkdenabled optical data center networks,'' {\em IEEE Internet of Things Journal}, 2022.

\bibitem{zhang2020flexible}
C. Zhang, Z. Liu, Y. Chen, J. Lu, and D. Liu, ``A flexible and generic gaussian sampler with power side-channel countermeasures for quantum-secure internet of things,'' {\em IEEE Internet of Things Journal}, vol. 7, no. 9, pp. 8167-8177, 2020.

\bibitem{m-xu-quantum-secured-space2022}
M. Xu, D. Niyato, Z. Xiong, J. Kang, X. Cao, X. S. Shen, and C. Miao, ``Quantum-Secured Space-Air-Ground Integrated Networks: Concept, Framework, and Case Study'', {\em arXiv:2204.08673}, 2022. 

\bibitem{b-chales-h-2014-quantum-cryp}
C. H. Bennett and G. Brassard, ``Quantum cryptography: Public key distribution and coin tossing,'' {\em Theoretical Computer Science}, vol. 560, pp. 7-11, 2014.

\bibitem{lo2012measurement}
H.-K. Lo, M. Curty, and B. Qi, ``Measurement-device-independent quantum key distribution,'' {\em Physical review letters}, vol. 108, no. 13, p. 130503, 2012.

\bibitem{y-cao-hybrid-trusted2021}
Y. Cao, Y. Zhao, J. Li, R. Lin, J. Zhang and J. Chen, ``Hybrid Trusted/Untrusted Relay-Based Quantum Key Distribution Over Optical Backbone Networks,'' in {\em IEEE Journal on Selected Areas in Communications}, vol. 39, no. 9, pp. 2701-2718, 2021.

\bibitem{kaewpuang2022resource}
R. Kaewpuang, M. Xu, D. Niyato, H. Yu, and Z. Xiong, ``Resource Allocation in Quantum Key Distribution (QKD) for Space-Air-Ground Integrated Networks'', {arXiv:2208.08009}, 2022. 

\bibitem{h-cui-space-air-ground2022}
H. Cui and {\em et al.}, ``Space-air-ground integrated network (SAGIN) for 6G: Requirements, architecture and challenges,'' in {\em China Communications}, vol. 19, no. 2, pp. 90-108, 2022.

\bibitem{s-k-liao-satellite-to-ground2017}
S.-K. Liao and {\em et al.}, ``Satellite-to-ground quantum key distribution,'' {\em Nature}, vol. 549, no. 7670, pp. 43-47, 2017.

\bibitem{j-yin-entanglement-based2020}
J. Yin and {\em et al.}, ``Entanglement-based secure quantum cryptography over 1,120 kilometres,'' {\em Nature}, vol. 582, no. 7813, pp. 501-505, 2020.

\bibitem{h-y-liu-optical-relayed2021}
H.-Y. Liu and {\em et al.}, ``Optical-relayed entanglement distribution using drones as mobile nodes,'' {\em Physical Review Letters}, vol. 126, no. 2, p. 020503, 2021.

\bibitem{h-y-liu-drone-based-entanglement2020}
H.-Y. Liu and {\em et al.}, ``Drone-based entanglement distribution towards mobile quantum networks,'' {\em National science review}, vol. 7, no. 5, pp. 921-928, 2020.

\bibitem{g-basak-semantic-comm-game2018}
B. G\"{u}ler, A. Yener, and A. Swami, ``The semantic communication game,'' {\em IEEE Transactions on Cognitive Communications and Networking}, vol. 4, no. 4, pp. 787-802, 2018.
\bibitem{a-wonfor-field-trail2019}
A. Wonfor, C. White, A. Bahrami, J. Pearse, G. Duan, A. Straw, T. Edwards, T. Spiller, R. Penty, and A. Lord, ``Field trial of multi-node, coherent-one-way quantum key distribution with encrypted 5x100G dwdm transmission system,'' in {\em 45th European Conference on Optical Communication (ECOC 2019)}, pp. 1-4, 2019.

\bibitem{h-xie-deep-learning2021}
H. Xie, Z. Qin, G. Y. Li and B. -H. Juang, ``Deep Learning Enabled Semantic Communication Systems,'' in {\em IEEE Transactions on Signal Processing}, vol. 69, pp. 2663-2675, 2021.

\bibitem{j-redmon-computer-vision2018}
J. Redmon and A. Farhadi, ``YOLOv3: An Incremental Improvement'', {\em arXiv:1804.02767}, 2018.

\bibitem{Brige1997}
J. R. Birge and F. Louveaux, {\em Introduction to stochastic programming}, {\em Springer Science \& Business Media}, 2011.

\bibitem{peleg2007introduction}
B. Peleg and P. Sudh\"{o}lter, {\em Introduction to the theory of cooperative games}, {\em Springer}, Nov. 2010.

\bibitem{Goyal2005}
S. Goyal and F. Vega-Redondo, ``Network formation and social coordination,'' { \em Games and Economic Behavior}, vol. 50, no. 2, pp. 178-207, Feb. 2005.

\bibitem{Gams}
GAMS, ``GAMS Solvers'', https://www.gams.com/. 


\end{thebibliography}



%XXXXXXXXXXXXXXXXXXXXXXXXXXXXXXXXXXXXXXXXXXXXXXXXXXXXXXXXXXXXXXXXXXXXXXXXXXXXXXXXXXXXXX%

%\bibitem{l-fatemeh-semantic-aware2022}
%F. Lotfi, O. Semiari, and W. Saad, ``Semantic-Aware Collaborative Deep Reinforcement Learning Over Wireless Cellular Networks,'' {\em ICC 2022 - IEEE International Conference on Communications}, pp. 5256-5261, 2022.


%\bibitem{g-stamatakis-semantic-aware2022}
%G. Stamatakis, N. Pappas, A. Fragkiadakis, and A. Traganitis, ``Semantics-Aware Active Fault Detection in Status Updating Systems'', {\em arXiv:2202.00923 }, 2022. 


%\bibitem{li2017distributed}
%R. Li, H. Asaeda, and J. Li, ``A Distributed Publisher-Driven Secure Data Sharing Scheme for Information-Centric IoT,'' in {\em IEEE Internet of Things Journal}, vol. 4, no. 3, pp. 791-803, 2017.

%XXXXXXXXXXXXXXXXXXXXXXXXXXXXXXXXXXXXXXXXXXXXXXXXXXXXXXXXXXXXXXXXXXXXXXXXXXXXXXXXXXXXXX%

%\bibitem{l-fatemeh-semantic-aware2022}
%F. Lotfi, O. Semiari, and W. Saad, ``Semantic-Aware Collaborative Deep Reinforcement Learning Over Wireless Cellular Networks,'' {\em ICC 2022 - IEEE International Conference on Communications}, pp. 5256-5261, 2022.

%\bibitem{g-stamatakis-semantic-aware2022}
%G. Stamatakis, N. Pappas, A. Fragkiadakis, and A. Traganitis, ``Semantics-Aware Active Fault Detection in Status Updating Systems'', {\em arXiv:2202.00923 }, 2022. 

%\bibitem{li2017distributed}
%R. Li, H. Asaeda, and J. Li, ``A Distributed Publisher-Driven Secure Data Sharing Scheme for Information-Centric IoT,'' in {\em IEEE Internet of Things Journal}, vol. 4, no. 3, pp. 791-803, 2017.

\end{document}